\documentclass[]{aa}
\usepackage{natbib}
\usepackage{graphicx}
\usepackage{txfonts}

\bibpunct{(}{)}{;}{a}{}{,}

\newcommand\bb[1] {   \mbox{\boldmath{$#1$}}  }
\newcommand\del{\bb{\nabla}}
\newcommand\bcdot{\bb{\cdot}}
\newcommand\btimes{\bb{\times}}

\begin{document}

\title{Global MHD simulations of stratified and turbulent protoplanetary
  discs. I. Model properties}
\author{ S\'ebastien Fromang \inst{1,2} and Richard P. Nelson \inst{2}}

\offprints{S.Fromang}

\institute{Department of Applied Mathematics
and Theoretical Physics, University of Cambridge, Centre for
Mathematical Sciences, Wilberforce Road, Cambridge, CB3 0WA, UK \and  Astronomy Unit, Queen Mary, University of London, 
Mile End Road, London E1 4NS \\ \email{S.Fromang@damtp.cam.ac.uk}}

\date{Accepted; Received; in original form;}

\label{firstpage}

\abstract
{}
{We present the results of global 3-D MHD simulations of stratified
and turbulent protoplanetary disc models. The aim of this work is to develop 
thin disc models capable of sustaining turbulence for long run times,
which can be used for on--going studies of planet formation in turbulent discs.}
{The results are obtained using two codes written in spherical
  coordinates: GLOBAL and NIRVANA. Both are time--explicit and use
  finite differences along with the Constrained Transport algorithm to
  evolve the equations of MHD.}
{In the presence of a weak toroidal magnetic field, a thin protoplanetary disc
  in hydrostatic equilibrium is destabilised by the 
  magnetorotational instability (MRI). When
  the resolution is large enough ($\sim 25$ vertical grid cells per scale height), the
  entire disc settles into a turbulent quasi steady--state after about
  $300$ orbits. Angular momentum is transported outward such that the
  standard $\alpha$ parameter is roughly $4-6 \times 10^{-3}$. We
  find that the initial toroidal flux is expelled from the disc
  midplane and that the disc behaves essentially as a quasi--zero net flux
  disc for the remainder of the simulation. As in previous studies,
  the disc develops a dual structure composed of an MRI--driven
  turbulent core around its midplane, and a magnetised corona stable
  to the MRI near its surface. By varying disc parameters and boundary 
  conditions, we show that these basic  properties of the models are robust.}
{The high resolution disc models we present in this paper achieve a quasi--steady state
 and sustain turbulence for hundreds of orbits. As such, they are ideally 
 suited to the study of outstanding problems in planet formation such as
 disc--planet interactions and dust dynamics.}

\authorrunning{S.Fromang \&  R.P.Nelson}
\titlerunning{Stratified protoplanetary discs models}
\maketitle

\section{Introduction}

Observational surveys of star forming regions in the Galaxy 
have revealed the ubiquity of rotationally supported discs of gas and
dust orbiting young stars 
\citep[e.g.][]{beckwith&sargent96,odelletal93,staufferetal94,
siciliaaguilaretal06,kessleretal06}.
It is commonly believed that these discs are the likely sites of
planetary formation \citep{safronov69,lissauer93}.
The discovery of numerous extrasolar planets
has increased the need for greater understanding of their properties 
so that accurate models of planet formation can be developed.

These ``protoplanetary'' discs often show evidence for active
accretion with a canonical
mass flow rate onto the central star of $\sim 10^{-8}$ M$_{\odot}$ yr$^{-1}$ 
\citep[e.g.][]{siciliaaguilaretal04}, requiring a source of anomalous
viscosity to transport angular momentum outward.
It has long been believed that this is provided by turbulence within the
disc \citep[e.g.][]{shakura&sunyaev73}.
So far only one mechanism has been shown to work reliably:
MHD turbulence generated by the magnetorotational instability (MRI)
\citep{balbus&hawley91,balbus&hawley91b}.
Given the cool and dense nature of protoplanetary discs, there are
questions about the global applicability of the MRI in such
environments as the ionisation fraction is low \citep{blaes&balbus94}.
Models suggest that protoplanetary discs are likely to have both
magnetically active zones, where the disc is turbulent, and adjacent
magnetically `dead-zones' where the flow is laminar
\citep[e.g.][]{gammie96,fromang02,ilgner&nelson06a}. In this paper we
focus on ideal MHD simulations of protoplanetary discs. We will
examine the dynamics of `dead--zones' in future work.

Non linear numerical simulations performed using the local
shearing box formalism
\citep[e.g.][]{balbus&hawley91b,hawleyetal96,brandenburgetal96} have
shown that the saturated non linear outcome of the MRI is MHD
turbulence having an effective viscous stress parameter $\alpha$ 
between $\sim 5 \times 10^{-3}$ and $\sim 0.1$, depending on the
magnetic field configuration. Outward angular momentum transport can thus occur
at the rate required to match observed accretion signatures onto T
Tauri stars \citep[see, for example,][who quote $\alpha \sim 0.01$ as
  being suggested by the observations]{hartmannetal98}.  Much of this
early simulation work was performed in discs
with no vertical stratification, and so was useful in determining the
nonlinear outcome of the MRI,
but did not provide insights into the global structure of these
discs in either the radial or vertical direction.
The question of their vertical structure 
was addressed by \citet{stoneetal96} and \citet{miller&stone00} who performed
shearing box simulations of vertically stratified discs.
A basic result to come out of these studies is that the discs evolve to
a structure consisting of a dense region around the
miplane where turbulence is driven by the MRI, sandwiched between a 
tenuous and magnetically dominated
corona which is highly dynamic, but stable against the MRI.

Global MHD simulations of turbulent discs
\citep[e.g.][]{armitage98,hawley00,hawley01,steinacker&pap02,pap&nelson03a}
confirm the basic picture provided by the local shearing box simulations.
These early global simulations considered the radial structure 
of turbulent disc models, but employed non stratified cylindrical disc models.
Recent work has been performed examining the dynamics
of global, vertically stratified turbulent disc models,
and has focussed on thick accretion tori around black holes,
using either the Paczynski--Witta potential
\citep{hawley00,hawley&krolic01,hawleyetal01} or simulating accretion
flows in the full Kerr metric \citep[e.g.][]{devilliers03}. 
The starting conditions for these models are usually thick,
constant angular momentum tori for which the gas is assumed to be
non--radiating. These quickly evolve into thick accretion discs
for which $H/R>0.1$.
To date, there have been no published simulations of vertically stratified
thin discs undergoing MHD turbulence (i.e. discs more akin to
protoplanetary discs). In part this is because of the increased
computational burden associated with simulating thin discs
due to the higher resolution requirements.

The aim of this paper is to present and analyse a suite of
MHD simulations of global, stratified, turbulent, protoplanetary disc 
models with $H/R \le 0.1$. We focus on the global structure
of the discs, and analyse how modifications to the
boundary conditions and disc thickness affect the results.
The longer term goal is to use these models
to examine a range of problems in planetary formation in the context
of turbulent, vertically stratified discs. As such
this paper is the first in a series. Future publications will
address problems such as: the evolution of dust in turbulent discs;
the orbital evolution of planetesimals and low mass planets;
gap formation and gas accretion by giant protoplanets; the dynamical
evolution of dead--zones. A number of previous studies have addressed
these issues using cylindrical models:
\citet{nelson&pap03b,papaloizouetal04,nelson&pap04b,nelson05}
studied gap formation and planet formation in
turbulent discs. They showed in
particular that type I migration becomes stochastic because of the
turbulent density fluctuations in the disc. \citet{fromang&nelson05}
also used cylindrical models to study the radial migration
of solid bodies due to gas drag.
They found rapid accumulation of meter size bodies in
anticyclonic vortices apparently resulting from the
turbulence. Another key problem in planet formation, dust settling 
in turbulent discs, has been studied recently using shearing box models
\citep{johansenetal05,fromang&pap06,turneretal06}. Because of the local
approach, these analyses had to ignore radial drift. All of these
issues are likely to be affected by the simultaneous treatment of
radial and vertical stratification that we consider in this
present work.

The plan of the paper is as follows: in section~\ref{basic-equations},
we present the basic equations, notations, and diagnostics we use in this
work. The set-up of our simulations is detailed in
section~\ref{simulations} and we present their results in
section~\ref{results}. We focus in particular on their global
structure, and sensitivity to
numerical issues such as resolution and boundary conditions. Finally,
in section~\ref{conclusion}, we summarise our results and 
highlight future improvements that will be added to the models.

\section{Basic equations}
\label{basic-equations}

The equations we seek to solve are the standard equations of MHD in a
frame rotating with the angular velocity $\bb{\Omega_{rot}}$.
In Gaussian units these may be
expressed as:
\begin{eqnarray}
\frac{\partial \rho}{\partial t} + \del \bcdot (\rho \bb{v}) =  0 \,\label{contg} , \\
 \frac{\partial \bb{v}}{\partial t} + ( \bb{v} \bcdot \del )
\bb{v} + 2 \bb{\Omega_{rot}} \btimes \bb{v} =  - \frac{1}{\rho} \del P - \nabla \Phi + \frac{1}{4\pi\rho} (\del \btimes \bb{B})
\btimes \bb{B} \,\label{mog} , \\
 \frac{\partial \bb{B}}{\partial t}  =  \del \btimes ( \bb{v} \btimes
\bb{B}- \eta \nabla \btimes \bb{B} ) \, \label{induct}.
\label{mhd_eq}
\end{eqnarray}
The symbols have their usual meanings:
$\rho$ is the density, $\bb{v}$ is the velocity, $P$ is the gas pressure,
$\bb{B}$ is the magnetic field, $\eta$ is the magnetic diffusivity,
and $\Phi$ is the sum of the gravitational and centrifugal potentials.
In the following discussion we will use two systems of coordinates:
cylindrical coordinates $(R,\phi,Z)$ and spherical coordinates $(r,\theta,\phi)$.
Using the later, $\Phi$ becomes
\begin{equation}
\Phi=-\frac{1}{2}\Omega_{rot}^2 r^2 \sin^2 \theta - \frac{GM}{r}
\end{equation}
where $G$ is the gravitational constant and $M$ the mass of the
central protostar. To complete equations (\ref{contg})--(\ref{mhd_eq}),
 we use a locally isothermal equation of state:
\begin{equation}
P=c(R)^2 \rho
\end{equation}
where $c(R)$ is the sound speed which is a fixed function of position.

\subsection{Averaged quantities and connection to viscous disc theory}
\label{averages}
When presenting the results of our calculations we make use
of a number of averaged quantities. For a quantity $Q$, we define the average
${\overline Q(R,t)}$ through
\begin{equation}
{\overline Q}(R,t) = \frac{\frac{1}{\Delta \phi} \int \int \rho Q(R,\phi,Z,t)
dz d\phi}
{\frac{1}{\Delta \phi} \int \int \rho dz d \phi}
\label{average}
\end{equation}
where $\Delta \phi = \phi_{max} - \phi_{min}$ is the size of the
azimuthal domain (which is less than $2 \pi$ in the disc models we
present later in this paper). The integrals are taken over the total
disc height and azimuth. We define the disc surface density through
\begin{equation}
\Sigma(R)={\frac{1}{\Delta \phi} \int \int \rho dz d \phi}
\label{sigma}
\end{equation}
Note that without loss of generality we can also include a time--average within
our averaging procedure. In this case the quantity
${\overline Q}(R,t)$ becomes:
\begin{equation}
{\overline Q}(R,t) = \frac{1}{2\Delta t} \int_{t-\Delta t}^{t+\Delta t} {\overline Q}(R,t) dt
\label{time-average}
\end{equation}
where $2\Delta t$ is the time interval over which the averaging
procedure is performed.

We now consider the connection between the turbulent disc models and standard
viscous disc theory
\citep[e.g.][]{shakura&sunyaev73,balbus&pap99,pap&nelson03a}. Averaging
the continuity equation~(\ref{contg}) gives 
\begin{equation}
\frac{\partial \Sigma}{\partial t} + \frac{1}{R} \frac{\partial}{\partial R}
\left(R \Sigma {\overline v_R} \right) = 0.
\label{av-cont}
\end{equation}
Consider the azimuthal component of the momentum equation~(\ref{mog}).
Multiplying by $R$ to give an equation for the conservation of angular
momentum about the $Z$ axis, and averaging over the disc height and azimuth gives:
\begin{eqnarray}
\frac{\partial}{\partial t} \left( \Sigma {\overline j} \right)
+ \frac{1}{R} \frac{\partial}{\partial R} \left( R \Sigma {\overline j} \,
{\overline v_R} \right) = & - &\frac{1}{R} \frac{\partial}{\partial R} \left(
R^2 \Sigma \overline{\delta v_{\phi} \delta v_R} \right) \nonumber \\
& + & \frac{1}{R} \frac{\partial}{\partial R} 
\left[ R^2 \Sigma 
\overline{ \left( \frac{B_{\phi} B_R}{4 \pi \rho} \right) } \right].
\label{av-ang-mom}
\end{eqnarray}
Here $j$ is the specific angular momentum, $R v_{\phi}$, and
we have used the relations
$v_R = {\overline v_R} + \delta v_R$ and 
$v_{\phi} = {\overline v_{\phi}} + \delta v_{\phi}$,
where $\delta v_R$ and $\delta v_{\phi}$ are the fluctuations
in the radial and azimuthal velocities.
Note that we have neglected terms due to the pressure gradient in
equation~(\ref{av-ang-mom}) as the transport of
angular momentum within the disc is primarily due to Reynolds and
Maxwell stresses. These are
given by the two terms on the right hand side of equation~(\ref{av-ang-mom}),
and we define the Reynolds stress by
\begin{equation}
{\overline T_R}= \Sigma \overline{\delta v_{\phi} \delta v_R}
\label{reynolds}
\end{equation}
and the Maxwell stress by
\begin{equation}
{\overline T_M}=\Sigma \overline{ \left( \frac{B_{\phi} B_R}{4 \pi \rho} \right) }.
\label{maxwell}
\end{equation}
We define the viscous stress parameter
$\alpha$ through
\begin{equation}
\alpha = \frac{{\overline T_R} - {\overline T_M}}{{\overline P}}
\label{alpha}
\end{equation}
where ${\overline P}$ is the averaged gas pressure.
In the discussion of our simulation results presented later in this
paper we refer to $\alpha_M$ and $\alpha_{R}$, which are the
Maxwell and Reynolds stress contributions to the total $\alpha$
value defined in equation~(\ref{alpha}). We also discuss the volume 
averages of $\alpha$,  $\alpha_M$ and $\alpha_R$ which we denote as
$\langle \alpha \rangle$,  $\langle \alpha_M \rangle$ and
 $\langle \alpha_R \rangle$.

\noindent Combining equations~(\ref{av-cont}) and (\ref{av-ang-mom})
gives:
\begin{equation}
\Sigma \frac{\partial {\overline j}}{\partial t} + \Sigma {\overline v_R} 
\frac{\partial {\overline j}}{\partial R} = -\frac{1}{R} 
\frac{\partial}{\partial R} 
\left(R^2 {\overline T_R} - R^2 {\overline T_M} \right) 
\end{equation}
Assuming that the averaged specific angular momentum ${\overline j}(R)$
is time independent, we obtain an equation for the averaged
radial velocity that is the equivalent of a similar 
expression obtained in standard viscous thin disc theory:
\begin{equation}
{\overline v_R} = -\frac{1}{R \Sigma} \left(\frac{\partial {\overline j}}{\partial R}
\right)^{-1} \frac{\partial}{\partial R} 
\left(R^2 {\overline T_R} - R^2 {\overline T_M} \right).
\label{v_R}
\end{equation}
By obtaining time averaged values of ${\overline T_R}(R)$, ${\overline
  T_M}(R)$, ${\overline j}(R)$, ${\overline v_R}(R)$ and $\Sigma(R)$
  from our numerical simulations, we are able to compare the value of
  ${\overline v_R}(R)$ obtained directly in the simulations
to that predicted by equation~(\ref{v_R}). Hence we can examine
how well our 3--D turbulent and stratified protoplanetary disc models are
described by thin disc theory, subject to an appropriate time average.
When calculating these averages in the simulations, which are performed using
spherical polar coordinates, we perform the vertical integration
by replacing $dz \rightarrow r \sin{\theta} d \theta$,
and integrate along the meridian at fixed $r$ and $\phi$.

\begin{table*}[t]\begin{center}\begin{tabular}{@{}lcccccccc}\hline\hline
Model & $c_0$ & Resolution & Azimutal extent & Vertical extent  & Inner radial BC & Outer radial BC & Vertical BC &
Code \\\hline\hline
S1a & 0.07 & $(272,90,126)$ & $\pi/4$ & 0.3 & Reflecting & Reflecting & Outflow & GLOBAL \\
S1b & 0.07 & $(272,90,126)$ & $\pi/4$ & 0.3 & Reflecting & Reflecting & Outflow & NIRAVANA \\
S2 & 0.07 & $(455,150,213)$ & $\pi/4$ & 0.3 & Reflecting & Reflecting & Outflow & GLOBAL \\
S3 & 0.07 & $(455,150,285)$ & $\pi/4$ & 0.4 & Reflecting & Reflecting & Outflow & GLOBAL \\
S4 & 0.07 & $(456,150,210)$ & $\pi/4$ & 0.3 & Reflecting & Reflecting & Periodic & NIRVANA \\ 
S5 & 0.1  & $(360,120,210)$ & $\pi/4$ & 0.43 & Outflow & Reflecting & Outflow & NIRVANA \\
C1 & 0.07 & $(455,150,40)$ & $\pi/4 $& 0.42 & Reflecting & Reflecting & Periodic & GLOBAL \\
C2 & 0.07 & $(260,152,44)$ & $\pi/2 $& 0.28 & Reflecting &
Reflecting & Periodic & GLOBAL \\
\hline\hline
\end{tabular}
\caption{Properties of the models described in this paper. The first
  column gives the model label, the second gives the sound speed at
  $r=r_0$. The third, fourth and fifth columns describe the resolution
  and the extent of the numerical box. The type of boundary conditions
  we used are given in columns 6 ,7 and 8, while the last column 
  indicates the code used to run that particular model. For the
  detailed set--up of model C2, see section~\ref{cyl_setup} or
  \citet{fromang&nelson05}.}
\label{model properties}
\end{center}
\end{table*}

\section{Numerical Simulations}
\label{simulations}

We used two codes to solve the MHD equations described in 
section~\ref{basic-equations}: GLOBAL~\citep{hawley&stone95} and NIRVANA
\citep{ziegler&yorke97}. GLOBAL, originally written in cylindrical geometry,
was modified to operate using spherical coordinates.
Both codes are time explicit, use
finite-differences to calculate  spatial derivatives and the
Method of Characteristics Constrained Transport (MOCCT) 
algorithm to evolve the magnetic field. They have been
widely tested in the past on a variety of different problems
relating to turbulent protoplanetary discs, 
making them particularly well suited to undertake the computationally
demanding problem of simulating stratified protoplanetary discs models.

\subsection{Stratified disc model set-up}
\label{setup}

At time $t=0$ we specify a spatial distribution for the hydrodynamic
variables that is as close as possible to a stratified thin
disc in hydrostatic equilibrium. Except for the vertical
stratification, the model properties are
very close to those of \citet{pap&nelson03a}. The mass density $\rho$
and angular velocity $\Omega$ are defined using:
\begin{eqnarray}
\rho&=&\rho_0 \left( \frac{R_0}{R} \right)^{3/2}
exp \left(-\frac{Z^2}{2H^2} \right) \, , \\
\Omega&=& \sqrt{\frac{GM}{R^3}}.
\end{eqnarray}
The radial and meridional velocities $v_r$ and $v_{\theta}$ are given small
random perturbations 
(using a uniform distribution with amplitude $2.5$ \% of the
  sound speed).
The disc semi-thickness $H$ is related to disc parameters through
\begin{equation}
H=\frac{c(R)}{\Omega}\, .
\end{equation}
In the absence of magnetic fields, the initial disc model 
is completely determined once the function
$c(R)$ is given. In the following, we used:
\begin{equation}
c(R)=c_0\left(\frac{R_0}{R}\right)^{1/2}
\end{equation} 
such that $H/R$ is constant in the models. Using these relations, the
surface density $\Sigma$ satifies
\begin{equation}
\Sigma=\Sigma_0 \left( \frac{R_0}{R} \right)^{1/2}
\end{equation}
where $\Sigma_0=2\rho_0 c_0 R_0^{3/2}/ \sqrt{GM}$.

The constants
appearing in the above equations are given values $GM=1$, $R_0=1$,
$\rho_0=1$ and $\Omega_{rot}=0.5$.
Depending on the models, $c_0$ is either equal to $0.07$
or $0.1$ (see table~\ref{model properties}). The computational domain
extends from $R_{in}=1$ to $R_{out}=8$ in the radial direction. The
vertical and azimuthal extent depend on the model. But in all of them,
at least $8.5$ scale heights are covered in $\theta$ in order to provide a good
description of both the disc midplane and corona. When discussing
simulation results in
this paper, time will be quoted in units of the orbital time at the
inner edge of the computational domain. With our definitions, a
time span of $500$ orbits at $r=R_{in}$ (which is the typical
duration of a given model) corresponds to $177$ orbits at $r=2$, $63$
orbits at $r=4$ and $27$ orbits at $r=7$.


Before adding the magnetic field, the above model 
was run as a purely
hydrodynamic disc. Using reflecting boundary conditions in
$r$, periodic boundary conditions in $\phi$ and outflow boundary
conditions in $\theta$, we found it to stay very close to the initial
model description presented above. In particular, we found no sign of
the transient growth of hydrodynamic perturbations recently proposed
in the literature
\citep{ioannou&kakouris01,afshordietal05,mukhopadhyayetal05}. The
kinetic energy then decays on a time scale of $\sim 10$ orbits (some
wavelike motions are then excited by the imperfect nature of the
boundary conditions on time scales of hundreds of orbits, but the
associated kinetic energy always remains smaller than in the MHD case
by at least an order of magnitude).

In the MHD runs, a toroidal magnetic field was added to the disc at
$t=0$. It is defined to be nonzero in the region of
the disc satisfying $2.5<r<6$ and
$|\theta-\pi/2|<\theta_{out}^{mag}$. We took $\theta_{out}^{mag}=0.1$
in all of the models, except for model S5 for which
$\theta_{out}^{mag}=0.2$. The strength of the
field is such that the ratio $\beta$ of the thermal pressure to the
magnetic pressure is everywhere equal to $\beta_0=25$. Previous global
models of such configurations have proved to be unstable to the MRI and
to generate MHD turbulence that spreads into the entire computational
domain. We emphasize here that the flux of this magnetic field in the $\phi$
direction is nonzero at the beginning of the calculation.

\subsubsection{Boundary conditions}

We now come onto the boundary conditions.
In principle it is possible to use reflecting or outflow boundary conditions
at the inner and outer disc radii. Reflecting boundary conditions are
unsuitable as waves excited by the turbulence then tend to rattle around
the computational domain in an unphysical manner.
Outflow conditions were tried in test calculations but proved to
be unsuitable because of excessive mass loss out of the computational
domain. 
Instead we added a non turbulent buffer zone
close to each radial boundary (for all runs except S5, see
below), the inner one extending
from $r=R_{in}$ to $r=2$, the outer one extending from
$r=7$ to $r=R_{out}$. 
In both buffer zones, we included
a linear viscosity $q^{lin}$ with coefficient $C_1$ \citep[see the
  definitions of][]{stone&norman92a}, to damp fluid motion, and 
resistivity, $\eta$, to create a region that is
stable to the MRI and therefore non turbulent near
the boundary. Both dissipation coefficients increase linearly
from the boundary of the buffer zones toward the boundary of the
computational domain. In all cases except model S3, their maximum value
at the inner boundary is $C_1=C_{1,in}=1$ and $\eta=\eta_{in}=4.9
\times 10^{-4}$, while they reach $C_1=C_{1,out}=5$ and
$\eta=\eta_{out}=10^{-3}$ at the outer boundary. In model S3, we used
$C_{1,in}=1$, $C_{1,out}=30$, $\eta_{in}=2 \times 10^{-4}$ and
$\eta_{out}=10^{-2}$. Using this set-up, we found that the turbulent
velocity fluctuations and magnetic stresses damp smoothly in the
buffer zones.

For very long integration times, we found that mass tends to
accumulate at the interface between the inner buffer zone and the
active disc. This is expected for such a closed boundary condition. To
try to solve this problem, we modified the inner boundary condition in
model S5 and used a `viscous outflow condition' (the outer radial
boundary is kept unchanged). The viscous outflow condition specifies the radial
outflow velocity in the inner radial boundary using the expression
$v_r=-3/2 \alpha c_s^2/\Omega$ there. A value of $\alpha=5 \times
10^{-3}$ was adopted, in basic agreement with the transport coefficient
resulting from the nonlinear evolution of the MRI (see below). 
The magnetic field boundary condition at the disc inner edge 
defines the field to be normal to the boundary with magnitude
defined by the $\nabla . {\bf B}=0$ condition \citep[e.g.][]{hawley00}.

The boundary conditions in the $\theta$ direction are also of some
importance in constructing stratified models. In local simulations
using the shearing box approximation, \citet{stoneetal96} reported
very little effect of the boundary conditions. Nevertheless, we
investigated their importance by running two types of boundary
conditions: outflow and periodic. The latter is less
physical but has the advantage of preserving the total flux of the
magnetic field and the vanishing value of its divergence. The former
was implemented in two different ways: in GLOBAL, a zero gradient
boundary condition was applied on all the variables, including the
magnetic field \citep{miller&stone00}. 
NIRVANA, on the other hand, forces the magnetic field
to become normal to the boundary and still satisfy the condition $\del
\bcdot \bb{B}=0$ \citep{hawley00}. 
The comparison between these different alternatives
gives an insight to their importance on the global structure of the
disc.

In the upper layers of the disc, strong magnetic fields tend to
develop during the simulation. Because of the low density there, the
associated Alfv\'en velocity becomes very large and the time step
consequently very small. To prevent this from happening, we used the
Alfv\'en speed limiter first introduced by \citet{miller&stone00}
whose effect is to prevent the Alfv\'en speed becoming significantly
larger than a
user-defined threshold $v_A^c$. In the simulations presented in this
paper, we used a uniform value of $v_A^c=0.7$. 

\subsection{Cylindrical models}
\label{cyl_setup}
In order to examine the effect of stratification, we 
ran a non stratified cylindrical model, labelled C1. 
The set--up was very similar to that for model S2.
The vertical computational domain covered $-0.21 \le Z \le 0.21$,
and the initial magnetic field was a zero--net flux toroidal
field defined by:
\begin{equation}
B_{\phi}=B_0 \cos \left( 6\pi\frac{R-R_{in}}{R_{out}-R_{in}} \right) \, .
\end{equation}
$B_0$ is calculated such that the average value of $<\beta>=25$.
We also consider another cylindrical disc run, C2, which
used a net flux toroidal field with $\beta=270$ and was described
in detail in \citet{fromang&nelson05}. Because of the vertical
periodic boundaries, the magnetic flux is conserved in these models.

\subsection{Model descriptions}

The detailed properties of the models we ran are described in table
\ref{model properties}. Column $1$ gives their label. Column $2$
indicates the value of the parameter $c_0$. Given that
$GM=1$ in our simulations, this is also equal to $H/R$.
The resolution, the size of the
computational box in the $\phi$ and $\theta$ directions are
respectively given in columns $3$, $4$ and $5$. The type of boundary
conditions we use at the inner and outer radial boundary 
and in the $\theta$ direction
are shown by columns $6$, $7$ and 8. Finally, the code we use for each
particular model is indicated by column $9$. Note that runs C1 and C2
are cylindrical disc models (see section \ref{cyl_setup}). 

\section{Results}
\label{results}

We now present the results of our simulations. Although differences in detail
are found when comparing our various runs, a common picture of the
early evolution of our models emerges. The magnetised regions of the initial 
discs become unstable to the MRI on the local dynamical time scale, in
broad agreement with linear theory. Dynamo action associated with the
MRI causes
the local field to amplify, and as turbulence develops the field
buoyantly rises from the regions near the midplane
toward the disc surface. Initially the stresses associated with the field
as it rises into these upper regions cause rapid transport of 
angular momentum there, allowing the magnetised fluid 
to spread rapidly through the disc in the radial direction
near the disc surface.
Contemporaneously the growth of the MRI throughout the magnetised core
of the models causes the disc to become globally turbulent on a
time scale of $\sim 100$ orbits. The end result is a disc whose
global structure consists of a magnetically subdominant core within
$\sim 2 H$ of the disc midplane which remains highly turbulent
due to continuing instability to the MRI, above and below which reside
a magnetised and highly dynamic corona which becomes magnetically dominant
near the disc surface and stable to the MRI.
Angular momentum transport and associated mass
flow through the disc allows magnetic field to diffuse into the
resistive regions near the radial boundaries, where the fluid
remains non turbulent because of the resistivity and viscosity there.

\subsection{Dependance on resolution}
\label{resolution_needs}

\begin{figure}
\begin{center}
\includegraphics[scale=0.5]{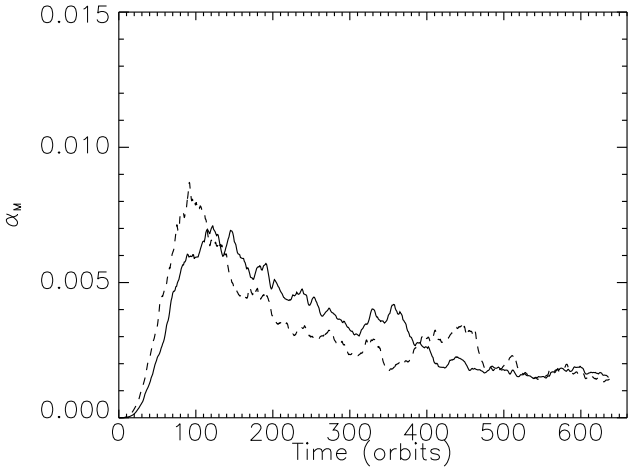}
\caption{Time history of $\langle \alpha_M \rangle $ 
    obtained in model S1a ({\it solid})
    and S1b ({\it dashed line}). The agreeement between the two
    curves shows that GLOBAL and NIRVANA produce similar
    results. However, in both runs, the rate of angular momentum
    transport displays signs that it decays with time, indicating
    that the resolution is too low in models S1a and S1b. At the end of
    both simulation, angular momentum transport is weak,
    with $\langle \alpha_M \rangle \sim 10^{-3}$ in both models.}
\label{compar_low_res}
\end{center}
\end{figure}
\begin{figure*}
\begin{center}
\includegraphics[scale=.6]{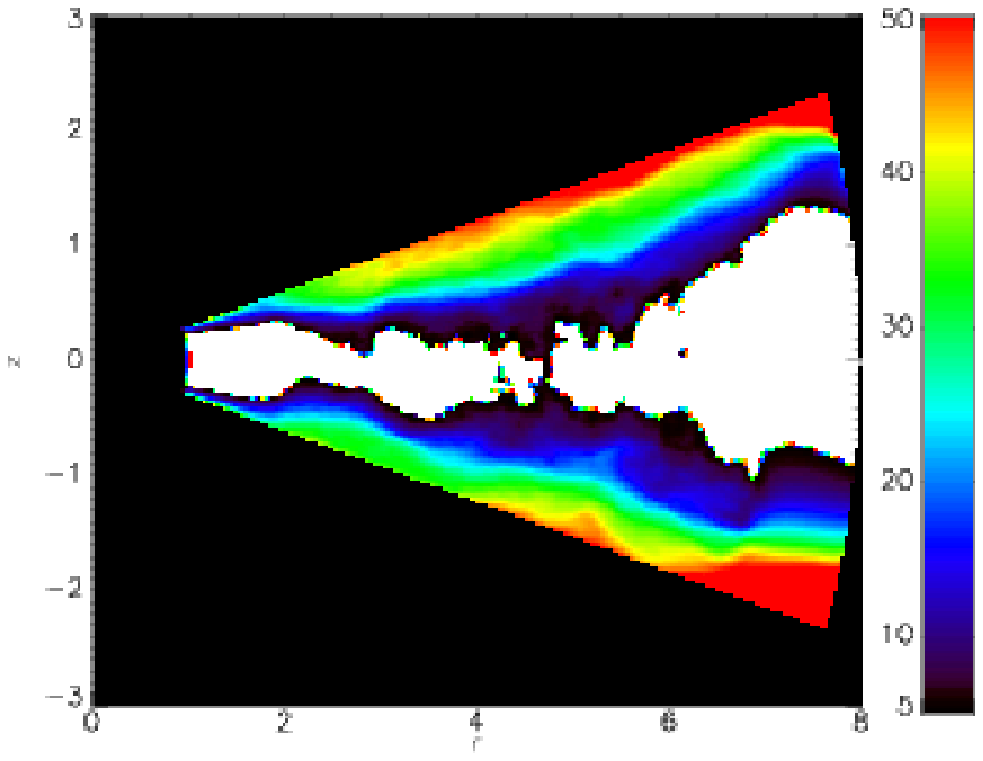}
\includegraphics[scale=0.71]{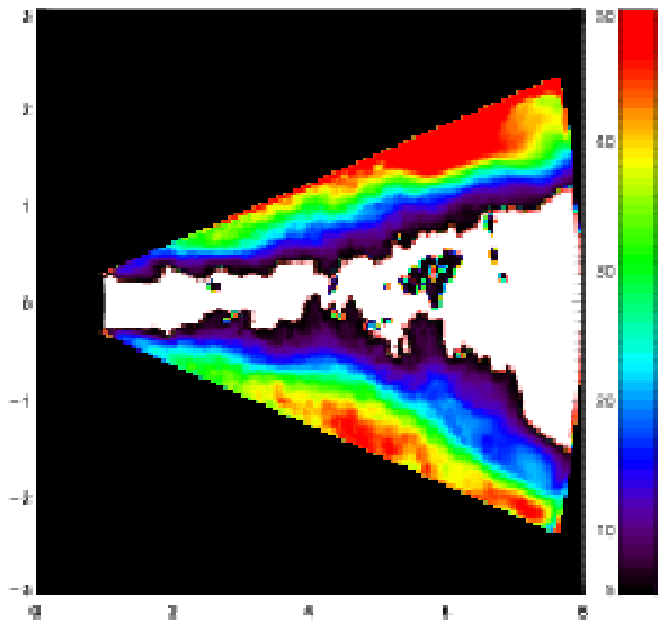}
\includegraphics[scale=0.6]{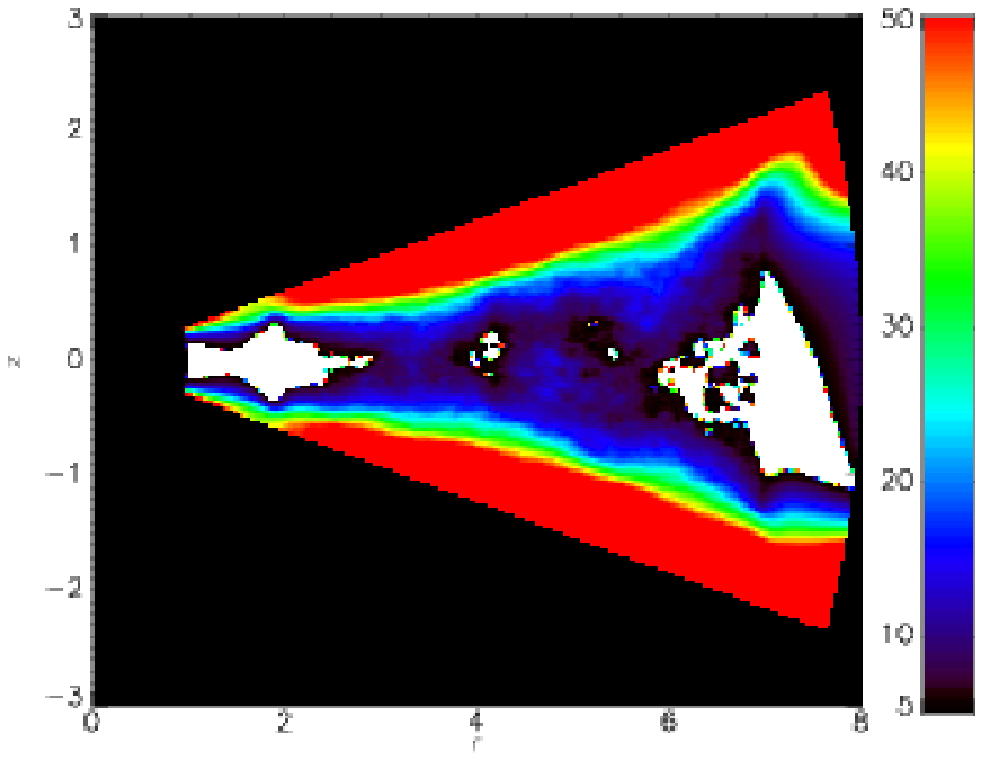}
\caption{The ratio of the wavelength of the most rapidly
growing MRI mode $\lambda_m$ to the local cell spacing $\Delta$ defined
in the text. From left to right, the different panels show results
from models S1a, S1b and S2. Any region coloured white has
$\lambda_m/\Delta < 5$, indicating that the fastest growing mode
of the MRI is not well resolved there.}
\label{resolve-fig}
\end{center}
\end{figure*}

The picture presented above is true for all our simulations in their
early phases. However, 
when designing useful and accurate stratified and turbulent
protoplanetary disc models, several issues need to be addressed before
presenting any detailed scientific results.

\begin{figure}
\begin{center}
\includegraphics[scale=0.5]{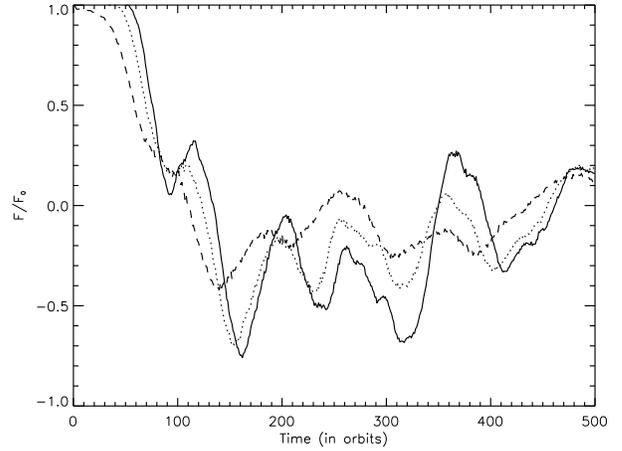}
\caption{Time history of the toroidal magnetic flux
  $F_{\phi}(\theta_0,t)$ threading the disc, normalized by its value at
  $t=0$, obtained in model S2. The solid line corresponds to
  $\theta_0=0.3$, the dotted line to $\theta_0=0.2$ and the dashed
  line to $\theta_0=0.1$. All of them show that the flux is expelled
  out of the midplane of the disc toward its corona.}
\label{flux_outflow}
\end{center}
\end{figure}

The duration of the models themselves is one of them. In global
cylindrical simulations, \citet{pap&nelson03a} demonstrated
that long integration times, over several hundred orbits are required
for the angular
momentum transport quantities to reach meaningful saturated
values. This is why, in this paper, we ran each model for at least $450$
orbits, and in some cases in excess of 600 orbits. 
In the following, we will show that this is sufficient to
reach a steady state in the underlying disc
structure, which is established after between $250$ to $500$ orbits.

A second issue is to determine the minimum resolution required to allow
such models to maintain turbulence over long run times. 
To do so, we present here the results of models S1a and
S1b. They were respectively run with GLOBAL and NIRVANA and have a 
resolution $(N_r,N_{\phi},N_{\theta})=(272,90,126)$,
corresponding to $15$ grid cells per scale height in the vertical
direction ($\simeq 8$ zones per scale height in the azimuthal direction,
$\simeq 3$ zones per scale height in the radial direction at the
disc inner edge and $\simeq 22$ zones per scale height at the disc outer
edge). Being identical in their set-up, these two models
are also useful as a direct comparison between the two codes. For both
cases, we find that the MRI grows leading to fully developed MHD turbulence
and outward angular momentum transport. 

The time history of $\langle \alpha_M \rangle$ is shown in
figure~\ref{compar_low_res} for model S1a ({\it solid line}) and S1b
({\it dashed line}). The two curves show very good agreement overall. 
This simply indicates that GLOBAL and NIRVANA give very similar
results for the same problem. It is therefore meaningful to compare the
results obtained by the two codes when they use different starting
conditions and physical parameters. However, the slow 
decrease of $\langle \alpha_M \rangle$ with time in both
models shows that MHD turbulence is getting weaker 
as the simulations proceed. No steady state seems to be reached
in these simulations, apparently because the 
resolution used in models S1a and
S1b is insufficient for vigorous MHD turbulence to be 
sustained in long runs. Indeed, in order to maintain turbulence, the
simulations must be capable of resolving the unstable modes of the
MRI. The critical wavelength  for instability is given by (Balbus \&
Hawley 1991):
\begin{equation}
\lambda_c = \frac{2 \pi}{\sqrt{3}} \frac{v_A}{\Omega}
\label{crit}
\end{equation}
and the wavelength of the fastest growing mode is
\begin{equation}
\lambda_m = 2 \pi \sqrt{\frac{15}{16}} \frac{v_A}{\Omega},
\label{max}
\end{equation} 
where $v_A$ is the Alfv\'en speed defined by
\begin{equation}
v_A=\sqrt{\frac{B^2}{4\pi \rho}} \, .
\label{v_A}
\end{equation}

A simulation must have at least 5 grid cells
per wavelength $\lambda_m$ for the fastest growing mode to be resolved
\citep[e.g.][]{hawleyetal95,miller&stone00}.
Figure~\ref{resolve-fig}
shows a contour plot in the ($r$, $\theta$) plane which
displays the ratio $\lambda_m/\Delta$ for models S1a, S1b and S2.
Here $\Delta$ is
the local cell spacing (the diagonal distance between cell vertices
along a path that passes through the cell centre).
$\lambda_m$ is calculated from azimuthally and temporally
averaged values of the magnetic field, $B$, and density $\rho$.
The time averages were performed between the 500$^{\rm th}$ and
600$^{\rm th}$ orbit of models S1a and S1b. Regions of figure~\ref{resolve-fig}
which are coloured white correspond to regions where $\lambda_m/\Delta \le 5$.
It is clear that 
the MRI modes with wavelengths between $\lambda_c$ and $\lambda_m$ are
not well resolved in these models around the midplane, leading to the weak
and declining angular momentum transport shown in 
figure~\ref{compar_low_res}.
\begin{figure*}
\begin{center}
\includegraphics[scale=1.25]{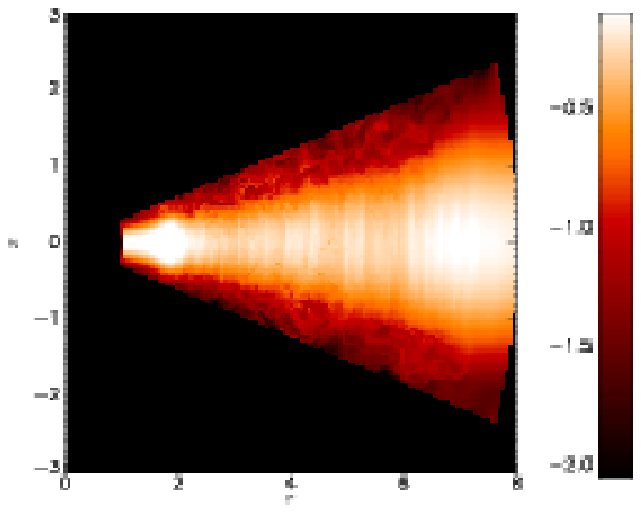}
\includegraphics[scale=1.25]{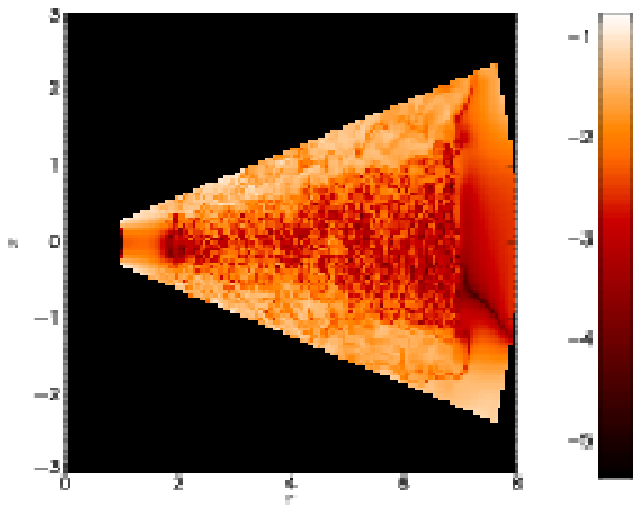}
\caption{The {\it left panel} shows the logarithm of the density distribution
in the ($r$, $\theta$) plane for model S2.
Before taking the logarithm we transformed $\rho$ using the following:
$\rho \rightarrow \rho \times r^{1.3} \times 
\exp{\left[Z^2/(2 \times 0.07^2)\right]} \times 
\exp{\left[-Z^2/(2 \times 0.1^2)\right]}$. This was done to increase the contrast 
in the figure.
The {\it right panel} shows the logarithm of 
the Alfv\'en speed in the $(r,\theta)$ plane, obtained in model S2 after 
$400$ orbits.}
\label{snapshots}
\end{center}
\end{figure*}

Motivated by these results, we increased the resolution by a factor
of $5/3$ to $25$ vertical grid
cells per scale height. This translates into a resolution
$(N_r,N_{\phi},N_{\theta})=(455,150,213)$. As we demonstrate in
section \ref{models2}, the results of this model,
labelled S2, suggest that a steady state is reached after
about $300$ orbits (see for example figure \ref{half_hist}) for this 
larger resolution. The rate of angular momentum transport seems to saturate,
giving a saturated value of 
$\langle \alpha \rangle \sim 4 \times 10^{-3}$  for
the remainder of the simulations. 
A contour plot of $\lambda_m/\Delta$ for model S2 
is shown in the right panel of figure~\ref{resolve-fig},
where $\lambda_m$ was time averaged between the 350$^{\rm th}$ and
450$^{\rm th}$ orbits. It is clear that $\lambda_m$ is well
resolved throughout the disc in this model, leading to the turbulence
being sustained. The value of $\lambda_m$ in the corona exceeds
the total disc model thickness of $\simeq 8.5 H$ for values of 
$|\theta - \pi/2| > 3 H/R$, such that the corona is stable 
against the MRI.
In the rest of this paper we will
describe results of models obtained using the same resolution as model S2.
Unfortunately, this leads to very long computing times, at
the limit of
present day capacities: on standard Pentium 2.8GHz Xeon chips each
model requires about 50000 CPU hours, or $\sim 6$ CPU years. Of course, 
it is still possible that MHD turbulence dies on time scales of several
hundreds of orbits even for this large resolution, but the limited
computational power available at the present time prevents running them
for longer than about $500$ orbits. It is probably also the case
that these simulations have not reached full numerical convergence.
Figure~\ref{resolve-fig} indicates that the smaller wavelength
unstable modes $\sim \lambda_c$ are at best marginally resolved in model
S2, such that a larger resolution may allow these modes to be more
active. Nevertheless, the saturated state
we generally obtained for the last $200$ to $300$ orbits sustains
turbulence well enough to
extract meaningful diagnostics describing the structure of the disc.

It is rather surprising that model S1a and S1b fail to show sustained
MHD turbulence, as the resolution used is similar to
that used by \citet{nelson05} and \citet{fromang&nelson05}, who
reported sustained turbulence in cylindrical disc models initiated with 
net flux toroidal magnetic fields. This is because the magnetic
field is trapped in the midplane of the disc in these cylindrical
models while it is expelled to the coronae and out of the
computational domain through the open boundaries in the stratified models
presented in this paper. To illustrate this result, we plot in
figure~\ref{flux_outflow} the time history of the toroidal magnetic
flux $F_{\phi}$ as obtained in model S1a. It is defined as
\begin{equation}
F_{\phi}(\theta_0,t)=\int_{-\theta_0}^{\theta_0} \int_{R_{in}}^{R_{out}} B_{\phi}(r,0,\theta,t) r dr
d\theta \, ,
\label{flux}
\end{equation}
and in figure~\ref{flux_outflow} is normalized by its initial value
  $F_0=F_{\phi}(\theta_0,0)$. The different curves in this figure
  corresponds to $\theta_0=0.3$ ({\it solid
  line}), $\theta_0=0.2$ ({\it dotted line}) and $\theta_0=0.1$ ({\it
  dashed line}). All decrease from $1$ to small values during the
first $100$ orbits (which corresponds to the growth of the MRI 
throughout the whole disc) and
then oscillate around zero. 
This behavior shows
that the magnetic flux is gradually expelled from the midplane toward
the corona of the disc and out of the computational domain
(the delay shown by the solid line to decay adds
further weight to this conclusion). The oscillations between positive and
negative values indicate that the adopted (vertical) boundary conditions
allow azimuthal net flux into the domain, although essentially zero
poloidal net flux enters.
In the midplane of the disc, the amplitude
of the oscillations is about $20-30\%$.
After about $150$ orbits, the disc
midplane essentially behaves as if MHD turbulence had been initiated 
using a zero net flux toroidal magnetic field. This requires a larger
numerical resolution for the turbulence to be sustained over large
periods (see the discussion by \citeauthor{nelson05}
\citeyear{nelson05}) and explains why models S1a and S1b fail to reach a
saturated state while cylindrical models using an equivalent resolution
do.

\subsection{The fiducial run -- model S2}
\label{models2}

We describe in this section the general properties of model S2, which
was run for $500$ orbits. As described above, the MRI grows in the
first $100$ orbits, before developing into MHD turbulence, which then
diffuses over the entire computational box (except for the buffer
zones described in section~\ref{setup}). The disc settles into a 
quasi--steady state after $250$ orbits for the remainder of the
simulation. The structure of the disc after $400$ orbits is illustrated in
figure~\ref{snapshots}. The left hand snapshot shows the
distribution of the logarithm of the density in the $(r,\theta)$ plane
and the right hand panel represents the logarithm of the Alfv\'en
speed $v_A$.
Strong radially propagating density waves are seen in the
former, while the small scale structure of the Alfv\'en speed in the
latter confirms the turbulent nature of the flow,
and also shows that the disc forms a structure consisting of
a turbulent, magnetically subdominant `core' near the midplane where 
the Alfv\'en speed is small,
and a magnetically dominant corona in the upper regions of the disc
where the Alfv\'en speed is large.

\subsubsection{Angular momentum transport}

\begin{figure}
\begin{center}
\includegraphics[scale=0.5]{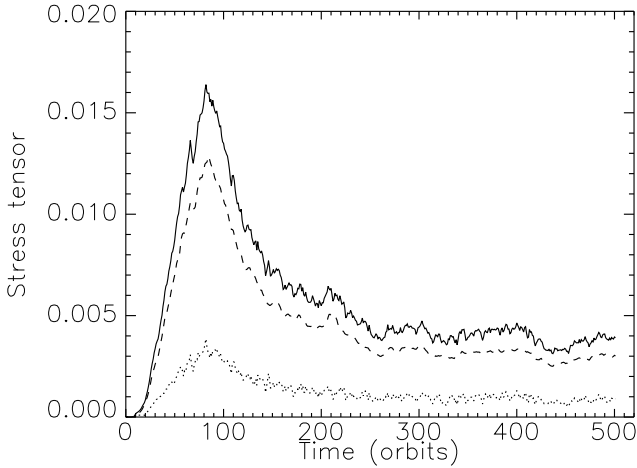}
\caption{Time history of the volume averaged value of 
  $\langle \alpha \rangle $
  ({\it solid line}), $\langle \alpha_M \rangle$ 
  ({\it dashed line}) and $\langle \alpha_R \rangle$
  ({\it dotted line}) in model S2. After an initial evolution driven
  by the linear growth of the MRI between $0$ and $250$ orbits, the
  disc settles into a quasi steady state for the remaining part of the
  simulation during which $\langle \alpha \rangle \sim 0.004$.}
\label{half_hist}
\end{center}
\end{figure}

\begin{figure}
\begin{center}
\includegraphics[scale=0.5]{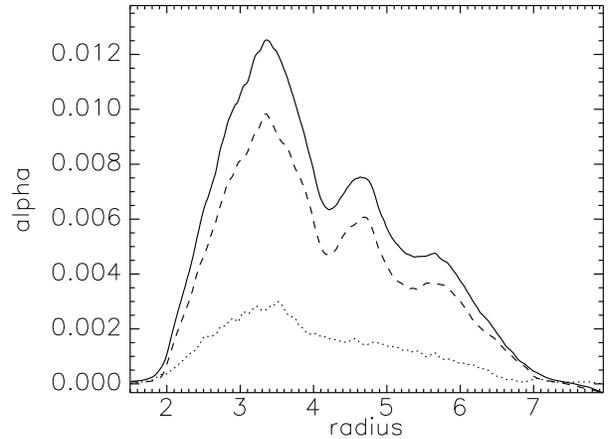}
\caption{Radial profile of $\alpha$
  ({\it solid line}), $\alpha_M$ ({\it dashed line}) and $\alpha_R$
  ({\it dotted line}) in model S2. The curves are time--averaged
  between $t=350$ and $t=450$ orbits.}
\label{alpha_mean}
\end{center}
\end{figure}

In this section, we quantify the rate of angular momentum transport
resulting from the MHD turbulence. It is measured by the sum of the
Maxwell and Reynolds stresses, already discussed in 
section~\ref{averages}.
The time history of $\langle \alpha \rangle $,
$\langle \alpha_M \rangle$ and $\langle \alpha_R \rangle$ 
are plotted in figure~\ref{half_hist}
respectively with the {\it solid, dashed} and {\it dotted}
lines. As reported
before for global MHD disc simulations
\citep{hawley00,hawley01,steinacker&pap02,pap&nelson03a}, the Maxwell
stress dominates over the Reynolds stress in the approximate ratio
of 3:1. The
rate of angular momentum transport saturates after $250$ orbits and
shows an almost constant value of 
$\langle \alpha \rangle \sim 4 \times 10^{-3}$ for
the remainder of the simulation. This is very similar to results
reported for zero--net flux numerical simulations of cylindrical discs
which further support the conclusion that our stratified disc model
behaves like a zero--net flux disc. \citet{pap&nelson03a} report values
of $\langle \alpha \rangle$ in the range 0.002 to 0.005 for a suite
of zero net flux cylindrical disc models, and model C1 achieves a 
saturated state with $\langle \alpha \rangle \simeq 0.002$.
This slightly smaller value of $\langle \alpha \rangle$ is
consistent with the smaller vertical resolution used in model C1 compared
with the models presented by \citet{pap&nelson03a}.
The radial profile of the Maxwell
and Reynolds stresses, normalized by $\bar{P}$, are represented in
figure~\ref{alpha_mean} respectively with the {\it dashed} and {\it dotted}
line. The {\it solid} line simply represents $\alpha$,
the sum of the
two. As reported by \citet{pap&nelson03a}, we found large fluctuations
in snapshots of these quantities. The smooth profile shown in
figure~\ref{alpha_mean} results from an averaging between $350$ and
$450$ orbits. We note that time averaged $\alpha$ profiles in
global MHD simulations (including these) often show variations between
a factor of 2 and 3 across the disc \citep[see
  e.g.][]{pap&nelson03a,steinacker&pap02}. The origin of these
variations is not clear, but may be related to the fact that
fluctuations in $\alpha$ anticorrelate with changes to the local
density and pressure due to mass transport. Local variations in
$\alpha$ may thus be reinforced. It is possible that these variations
will decrease with longer run times as the system loses memory of its
initial conditions.

\subsubsection{Structure of the corona}

\begin{figure}
\begin{center}
\includegraphics[scale=0.5]{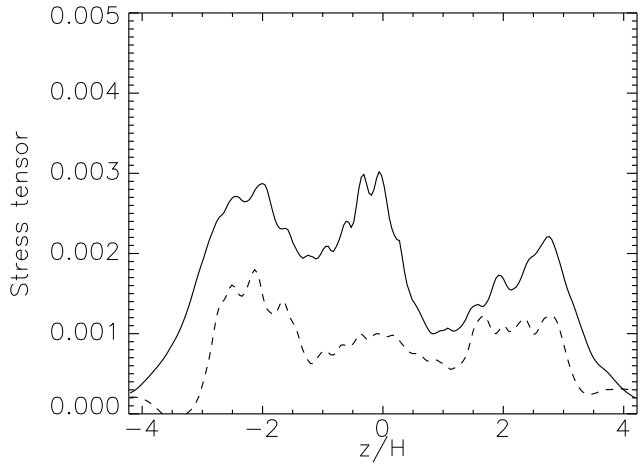}
\caption{Vertical profile of the Maxwell stress ({\it solid line}) and
the Reynolds stress ({\it dashed line}) obtained in model S2. The
results are time averaged between $350$ and $450$ orbits and in radius
between $r=4$ and $r=5$. Both curves are normalized by the midplane
pressure. A decrease of both stresses is apparent for $Z/H>3$.}
\label{maxwell_vert}
\end{center}
\end{figure}

\begin{figure}
\begin{center}
\includegraphics[scale=0.5]{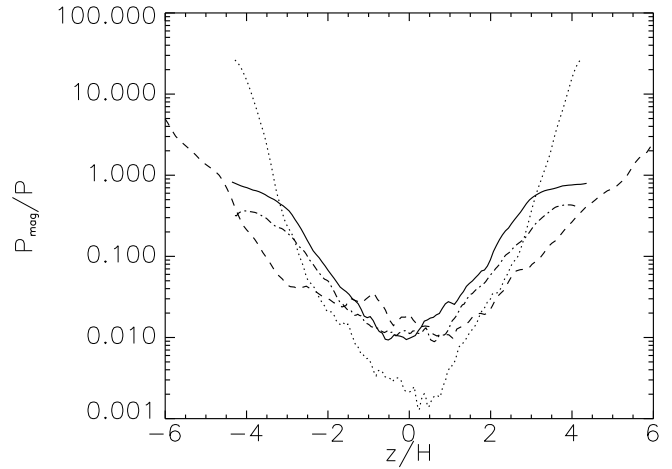}
\caption{Vertical profile of the radio $P_{mag}/P$ at $r=3.5$ obtained in model
  S2 ({\it solid line}, averaged between $350$ and $450$ orbits), model
  S3 ({\it dashed line}, averaged between $270$ and $310$ orbits),
  model S4 ({\it dotted line}, averaged between $500$ and $600$) and
  model S5 ({\it dotted--dashed line}, averaged between $500$ and
  $600$ orbits). For all models, the disc is composed of a weakly
  magnetised core and a corona where the field is close to
  equipartition. The differences between the different curves results
  from the boundary conditions and are further discussed in the text.}
\label{mag_press}
\end{center}
\end{figure}

\begin{figure}
\begin{center}
\includegraphics[scale=0.5]{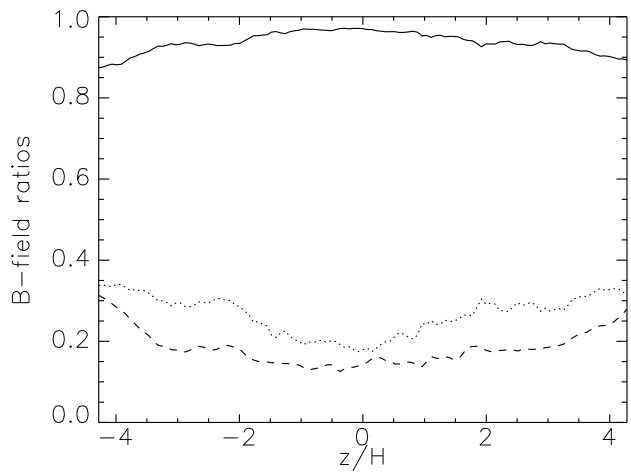}
\caption{The time averaged values of $|B_r|/B$, $|B_{\theta}|/B$ and
$|B_{\phi}|/B$ as a function of $\theta/(H/R)$ for model S5.
This was obtained by time averaging for 100 orbits at radius $r=3.5$.
The $\phi$ component of $B$ is represented by the {\it solid} line,
the $r$ component by the {\it dotted} line, and the $\theta$ component
by the {\it dashed} line.}
\label{bfield_ratio_s2}
\end{center}
\end{figure}

The results described above regarding the volume integrated properties of
angular momentum transport show little difference 
compared with zero--net flux
cylindrical disc models. In this section, we detail the structure of
the upper layers of the disc.

Figure~\ref{maxwell_vert} shows the vertical profile of the Maxwell
({\it solid line}) and Reynolds ({\it dashed line}) stresses,
normalised by the midplane pressure. Both curves are averaged in space
between $r=4$ and $r=5$ and in time between $t=350$ and $t=450$
orbits. In agreement with the results obtained in local disc
simulations \citep{miller&stone00}, both are fairly constant
for $|Z|<2.5H$ before decreasing in the upper layers of the disc. This
change is due to the establishment of a strongly magnetised
corona. This is illustrated by figure~\ref{mag_press}, which shows the
vertical profile of the ratio $P_{mag}/P$ at $r=3.5$ 
(the {\it solid} line corresponds to model S2). Simulation data were
averaged in time between
$350$ and $450$ orbits to smooth out the turbulent fluctuations (the
other curves represented in  figure~\ref{mag_press} plot the results
of models S3, S4 and S5 and will be discussed later). The 
relative strength of the magnetic field increases with $Z$, approaching
equipartition in the neighborhood of the lower and upper
boundaries. Even though the flow in the corona is not unstable to the
MRI, we found it to be highly dynamic, exhibiting strongly
time--dependent behaviour. Nevertheless, the structure of 
the corona is quite different to that in the midplane, with much
smaller scale magnetic field fluctuations near the
equatorial plane than in the corona of the disc. To try
to quantify the topology of the field in the disc,
figure~\ref{bfield_ratio_s2} shows the variation of $|B_r|/B$,
$|B_{\phi}|/B$, and $|B_{\theta}|/B$ versus $Z/H$, where
$B=\sqrt{B_r^2+B_{\theta}^2+B_{\phi}^2}$. This plot corresponds to a 
radial location $r=3.5$ and was obtained by time averaging for 100
orbits between $t=350$ and $450$ orbits.
It shows that the magnetic field topology is dominated by
the azimuthal component of the field, as expected because of the
strong Keplerian shear. However, as one moves away from the midplane
toward the disc surface, there are some indications that the $r$ and
$\theta$ component of the field become more important as their
relative magnitude increases by about a factor of two. 
We comment here that the boundary conditions imposed on the
magnetic field at the upper and lower disc surfaces were
zero gradient outflow conditions.

\subsubsection{Velocity and density fluctuations}

\begin{figure}
\begin{center}
\includegraphics[scale=0.5]{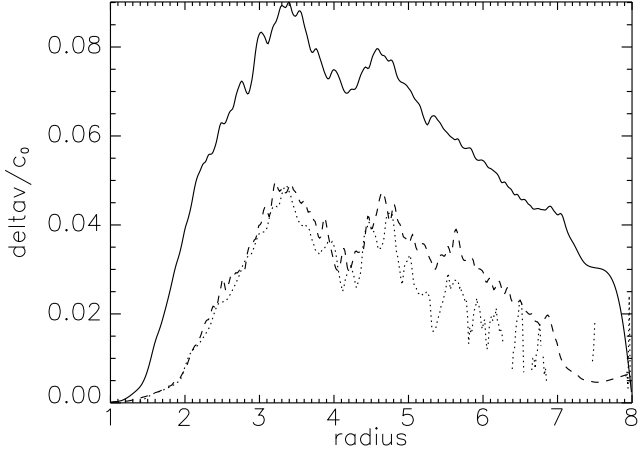}
\caption{Radial profile of the velocity fluctuations obtained in model
  S2. The curves, time--averaged between $t=350$ and $t=450$ orbits,
  correspond to radial ({\it solid line}), azimutal ({\it dashed
  line}) and vertical ({\it dotted line}) velocity fluctuations.}
\label{vel_fluc_radial}
\end{center}
\end{figure}

\begin{figure}
\begin{center}
\includegraphics[scale=0.5]{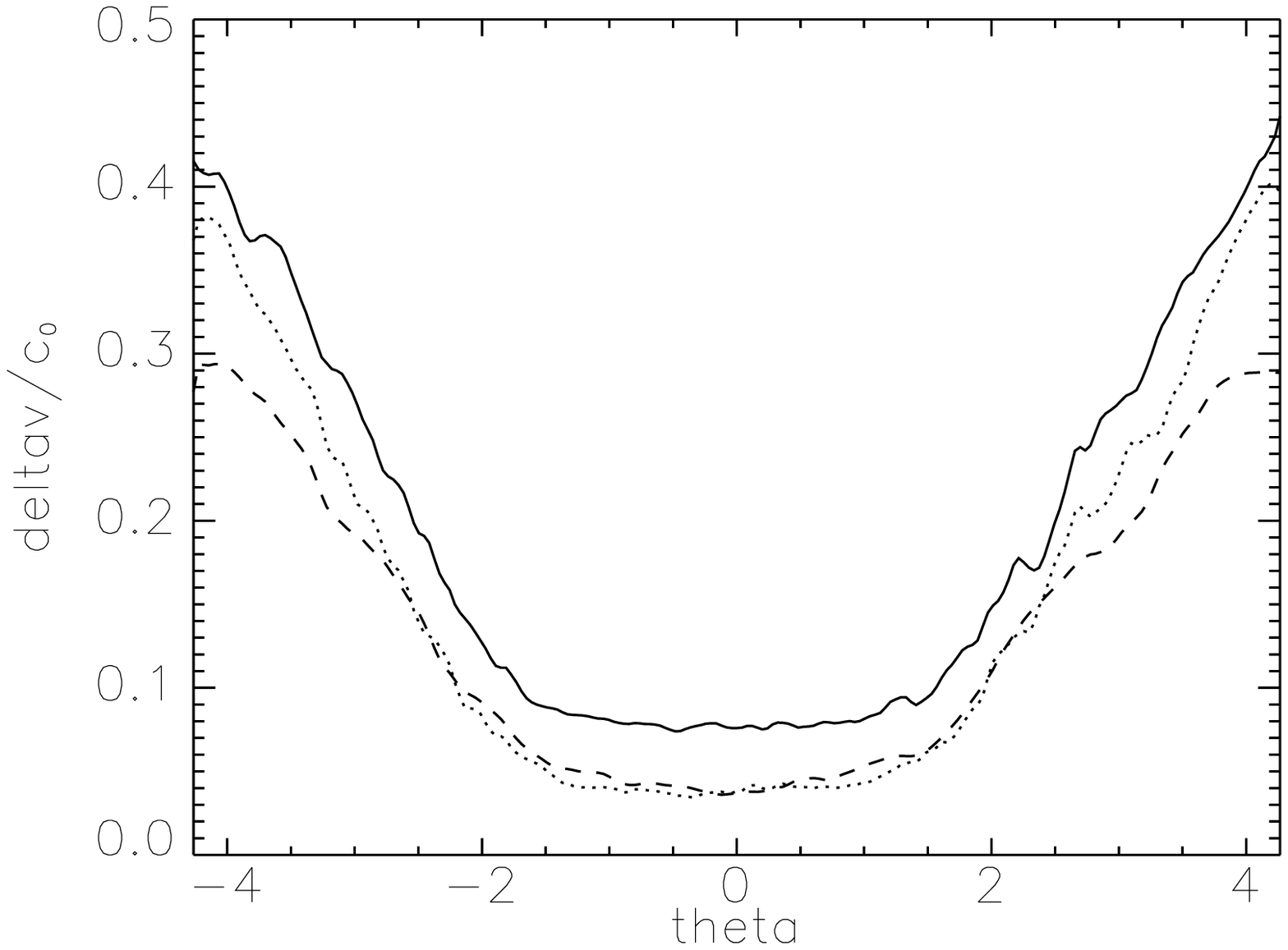}
\caption{Vertical profile of the velocity fluctuations obtained in model
  S2 at $r=4.5$. The curves, time--averaged between $t=350$ and $t=450$ orbits,
  correspond to radial ({\it solid line}), azimutal ({\it dashed
  line}) and vertical ({\it dotted line}) velocity fluctuations.}
\label{vel_fluc_vert}
\end{center}
\end{figure}

\begin{figure}
\begin{center}
\includegraphics[scale=1.25]{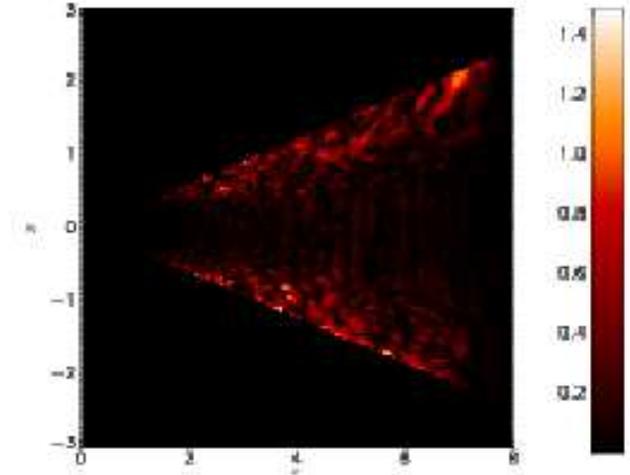}
\caption{Snapshots of the Mach number $M_s$ in the $(r,\theta)$ plane at
  $t=450$ orbits obtained in model S2. Motions in the bulk of the disc
  are largely subsonic while weak shocks (with $M_s$ between
  $1$ and $1.5$) can be seen in its corona.}
\label{mac_number}
\end{center}
\end{figure}

It is of interest to look at the velocity and density 
fluctuations generated  by
the turbulence in these models. Within the context of
planetary formation, they affect the
evolution of dust particles \citep{fromang&pap06}, whose spatial
distribution is important for the observational properties of
protoplanetary discs, as well as the orbits of larger bodies such
as boulders, planetesimals and protoplanets
\citep{nelson&pap04b,nelson05,fromang&nelson05}. 

\begin{figure}
\begin{center}
\includegraphics[scale=0.5]{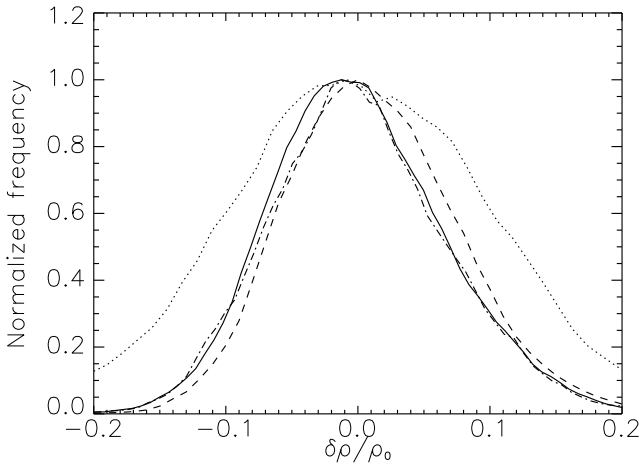}
\caption{Normalised frequency distribution of the relative density
  fluctuations in models S2 ({\it solid line}),
  S5 ({\it dashed line}), C1 ({\it dot--dashed
  line}) and C2 ({\it dotted line}). All curves are averaged in time
  for 100 orbits.
  Results obtained in run S2 and S5 are averaged in the volume
  $|Z|<H$ and $3<r<5$. Results obtained for runs C1 and C2 are averaged over
  the entire vertical extend and in radius between $r=3$ and
  $r=5$ for model C1, and between $r=2.5$ and $r=4$ for model C2. The
  curves for runs S2, S5 and C1 have a similar width, while model C2
  produces a wider distribution because it contained a net flux
  magnetic field.}
\label{dens_fluc}
\end{center}
\end{figure}

Figure~\ref{vel_fluc_radial} shows the radial profile of the velocity
perturbations obtained in
model S2, averaged between $t=350$ and $t=450$ orbits. The {\it solid,}
{\it dashed} and {\it dotted} line respectively represents the radial,
azimuthal and vertical velocity perturbation, normalized by the local sound
speed and averaged in space within $H/2$ around the disc
midplane. Typically, these fluctuations are all of the order of a few
percents of the sound speed. This is slightly smaller than previous results
obtained in the shearing box \citep{stoneetal96,fromang&pap06}, who
found values of the order of $10\%$. There is a marked tendency
for the radial fluctuations to be larger, by about a factor of two,
than the azimutal and vertical fluctuations. This is due to the
presence of density waves travelling radially through the disc. 
This is also
seen in local boxes but its effect is less pronounced in that case. 
The vertical profiles
of the fluctuations are shown, using the same conventions, in
figure~\ref{vel_fluc_vert}. As noted by \citet{miller&stone00} and
\citet{turneretal06} who presented
shearing box calculations, they increase in the upper layers of the
disc, where the averaged Mach number reaches $0.4$. This increase in
perturbed velocities arises because disturbances
that are excited near the high density midplane increase in
amplitude as they propagate vertically into the lower density regions,
due to conservation of wave action. In addition, the increasing
influence of the magnetic field with height means that propogating Alfv\'en 
waves in the disc corona can excite approximately sonic motions where
the Alfv\'en speed exceeds the sound speed. The result is that shocks are
generated in the upper regions of the disc, as illustrated
by figure~\ref{mac_number} which shows a snapshot in the ($r$,
$\theta$) plane of the perturbed velocity divided by the sound
speed. Weak shocks, for which $M_s$ reaches $1.5$, are visible on this image.

The normalised frequency distribution for the density fluctuations is
shown in figure~\ref{dens_fluc} for model S2 ({\it solid line}). The
results of the simulations were averaged in time between $350$ and
$450$ orbits in order to smooth out the turbulent fluctuations. An
additional spatial averaging was performed in the volume $3<r<5$ and
$|Z|<H$. The results show that the distribution is approximately
Gaussian with standard deviation $\sigma_{\rho} \simeq 0.08$. This
result agrees well with those of models S5 ({\it dashed line}) and
C1 ({\it dot--dashed line}), showing that
neither the boundary conditions nor the stratification have a strong
impact on the amplitude of the turbulent density
fluctuations near the midplane. However, the results of 
model C2, shown with the {\it dotted}
line, produce a broader distribution ($\sigma_{\rho} \simeq
0.13$). This is because the toroidal net flux is nonzero in this
model, while it vanishes in model S2, S5 and C1. This results in a
stronger MHD turbulence, along with slightly larger density
fluctuations. It is already
known that these density fluctuations may be responsible for
driving stochastic migration of low mass planets and planetesimals
\citep[e.g.][]{nelson&pap04b,nelson05}, and this effect
will be explored in more detail in a forthcoming paper within
the context of stratified, turbulent disc models.

\subsection{The effect of the boundary conditions in $\theta$}
We now present the results of simulations designed to test the
role of the vertical boundary conditions we have adopted.
We begin by examining a model whose vertical
domain is larger than model S2 but otherwise similar, and then
describe a model with periodic boundary conditions in
the vertical direction.

\subsubsection{A larger vertical domain -- model S3}

\begin{figure}
\begin{center}
\includegraphics[scale=0.5]{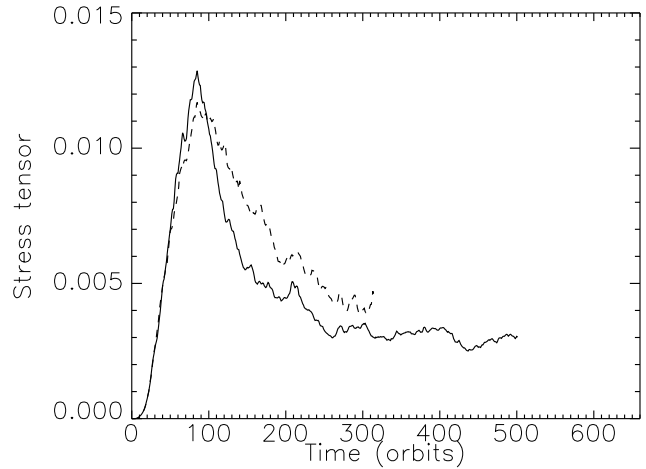}
\caption{Time history of the Maxwell stress in model S2 ({\it solid
  line}) and S3 ({\it solid line}). The latter has a larger
  vertical extent. The good agreement between the two curves shows
  that the boundary conditions and vertical extent
  have little effect on the results of
  model S2.}
\label{compar_big}
\end{center}
\end{figure}

We first focus on model S3. Compared to model S2 described above, it
has an extended domain in $\theta$: $|\theta - \pi/2| \le 0.4$. This
corresponds to 
a total of $5.7$ scale heights on both side of the equatorial
plane. For the actual resolution to remain the same, the total number
of grid cells was increased to
$(N_r,N_{\phi},N_{\theta})=(455,150,285)$. All the other parameters of
model S3 are otherwise identical to those of model S2. Note however
that the larger number of cells increased the computing time of model
S3. As a result it was run for only $325$ orbits.

Figure~\ref{compar_big} compares the time history of 
$\langle \alpha_M \rangle$ in
model S2 ({\it solid line}) and in model S3 ({\it dashed
  line}). Both curves show similar evolution. Moreover, model S3
starts to saturate after about $270$ orbits at a level that is
comparable to that of model S2. Figure~\ref{mag_press} also shows the
vertical profile of $P_{mag}/P$ ({\it dotted line}), averaged between
$270$ and $310$ orbits. It shows that model
S2 and S3 agree very well in the bulk of the disc.

These results demonstrate that the structure of the discs
described so far are not affected strongly by the
boundary conditions and vertical extent of the computational domain.

\begin{figure}
\begin{center}
\includegraphics[scale=0.5]{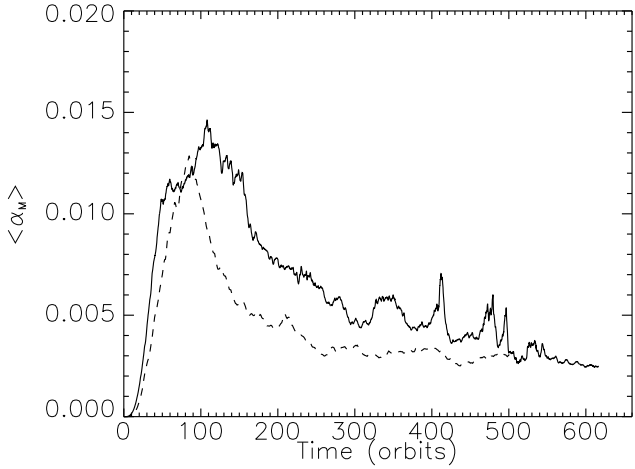}
\caption{Time evolution of the volume averaged
stress parameter $\langle \alpha_M \rangle$ for model S4
({\it solid line}) and S2
({\it dashed line}). Despite the different boundary conditions, the
two models display similar time history for the angular momentum
transport properties.}
\label{alpha_v_time_S4}
\end{center}
\end{figure}

\begin{figure}
\begin{center}
\includegraphics[scale=0.5]{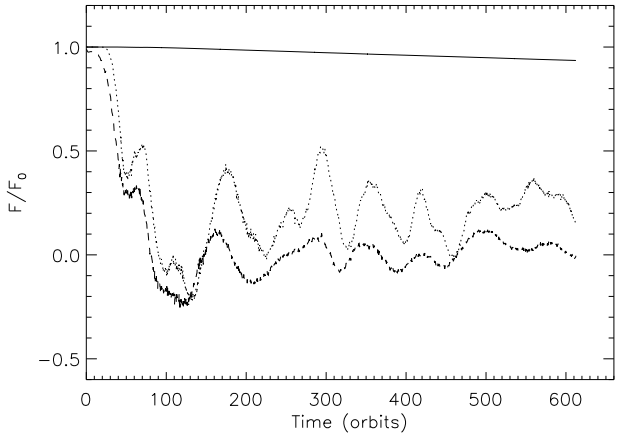}
\caption{The time evolution of the azimuthal magnetic flux for model S4,
normalised to its initial value.
The {\it dashed} line shows the total flux within an angular
 distance $|\theta - \pi/2 |\le H/R$ above and below the midplane.
The {\it dotted} line shows the flux within an angular distance
$| \theta - \pi/2| \le 2H/R$ above and below the midplane.
The {\it solid} line shows the flux throughout the whole disc,
and varies with time due to the imperfect nature of the
periodic boundary conditions adopted in the meridional direction.}
\label{magflux_S4}
\end{center}
\end{figure}

\subsubsection{Periodic boundary conditions -- model S4}
We have computed one model, run S4, which uses periodic boundary conditions
in the vertical direction (i.e. $\theta$--direction).
Clearly this is not a realistic boundary condition to use 
for an accretion disc, but it nonetheless provides a test of the role
that the vertical boundaries play in these simulations.
Before describing the results of the simulation in detail, we first
comment that the periodicity condition implemented in the NIRVANA
(or GLOBAL) code is imperfect when applied in the meridional direction.
This is because the physical sizes of the cells that overlap
at the top and bottom of the disc are different. The effect of this
is small, but one manifestation is that magnetic flux is not conserved
as it passes through the upper surface of the disc and re-enters
through the lower surface (and {\it vice versa}). 
As shown below, this causes a 5 \% decrease
in magnetic flux in the domain  over the simulation run time.
\begin{figure}
\begin{center}
\includegraphics[scale=1.2]{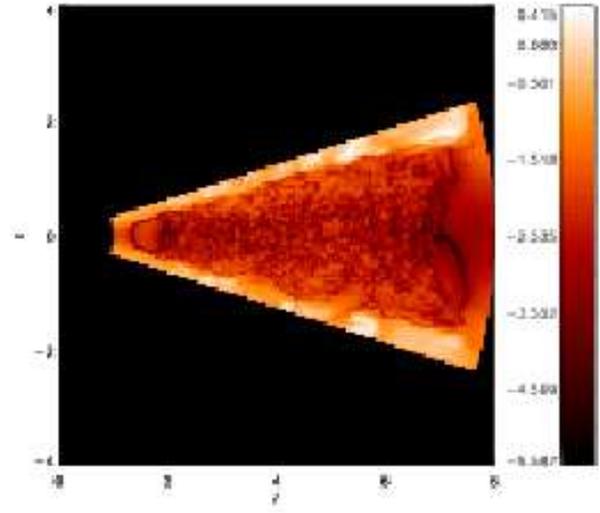}
\caption{A slice plotted in the ($r$, $\theta$) plane of the disc
showing the logarithm of the Alfv\'en speed for model S4.
Comparing with similar plots for models S2 (figure~\ref{snapshots})
and S5 (figure~\ref{alfven_S5}) shows that 
the trapping of magnetic flux in the upper disc regions of model S4 leads to
an elevated value of the Alfv\'en speed there.}
\label{alfven_S4}
\end{center}
\end{figure}

The time evolution of the volume averaged 
$\langle \alpha_M \rangle$ value is 
shown in figure~\ref{alpha_v_time_S4} for model S4 ({\it solid line})
and compared to model S2 ({\it dashed line}). The early rise and fall
of this quantity is similar in both models. One feature of the runs that 
appears throughout this paper is that the time required for
saturation of the turbulence is longer for the NIRVANA
simulations than for the GLOBAL ones. Nonetheless the final
states that are reached are very similar in each case.
Inspection of figure~\ref{alpha_v_time_S4} shows that during the
time interval 350--550 orbits, short duration outbursts are
observed in the $\alpha_M$ values. This phenomenon seems to be
explained by figure~\ref{magflux_S4}, which shows the
time evolution of the azimuthal magnetic flux in the
computational domain defined by equation~(\ref{flux}).
The {\it solid} line shows the total flux in the domain, and because of
the periodicity of the vertical boundaries this is approximately
conserved (the small non conservation was explained above).
The {\it dashed} and {\it dotted} lines show the magnetic flux
in the disc below heights $|\theta - \pi/2| < H/R$ and $|\theta -
\pi/2| < 2H/R$, respectively. As described for model S2, the onset of
the MRI and turbulence causes the magnetic flux to rise up through the disc
to form a magnetically dominated corona. In model S4 this magnetic 
flux is effectively trappped in the disc and is able to build up
to large values in the corona. Figure~\ref{magflux_S4}
shows that substantial flux occassionally comes down from
the corona into the turbulent core, within $\sim 2H/R$
of the midplane, where its presence causes an episodic increase in
turbulent activity.

The effects of trapping the magnetic flux in the domain on the evolution of
the disc is illustrated by figure~\ref{alfven_S4} which shows a snapshot
of the Alfv\'en speed plotted as a slice in the ($r$, $\theta$) plane.
The contrast in the Alfv\'en speed between the inner disc core and the 
upper corona is much greater in this case than was observed for model S2
because the magnetic field in the upper regions is larger.
This is further illustrated by figure~\ref{mag_press} which shows 
the ratio of magnetic to thermal pressure $P_{mag}/P$ as a function of
disc height for model S4 using the {\it dotted} line. 
This plot corresponds to $r=3.5$ and was obtained by
time averaging between $t=500$ and $t=600$ orbits.
It is clear that the corona in this case is highly dominated
by the magnetic field, with $P_{mag}/P \simeq 30$
near the disc surface, in contrast to the value of
$P_{mag}/P \simeq 1$ obtained in model S2. 

\begin{figure}
\begin{center}
\includegraphics[scale=1.2]{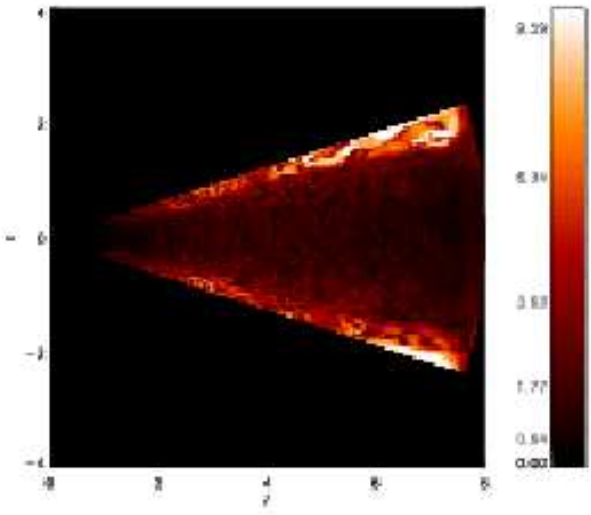}
\caption{Snapshot of velocity fluctuations in the disc normalised to the 
local sound speed, plotted in the ($r$, $\theta$) plane.
It is clear that strong shocks (Mach numbers $>10$ )
are generated in the upper disc regions due to the large magnetic forces there.}
\label{dv_S4}
\end{center}
\end{figure}

One effect of these large magnetic field and Alfv\'en speed values is that
large supersonic motions are generated in the disc corona.
This is illustrated by figure~\ref{dv_S4} which shows a snapshot
of the velocity perturbation divided by the local sound speed,
projected onto the ($r$, $\theta$) plane. Strong shocks can be seen
in the upper disc surface with Mach numbers $M_s$ in excess of 10 being common.
This is in contrast to the more quiescent state of the corona
obtained in model S2 where Mach numbers between 1 -- 2 are
observed. We note that $M_s \sim 10$ indicates that the maximum gas
  velocities in the corona are basically given by the Alfven speed
  limiter $v_A^c$
  (which is roughly ten times the sound speed). This
  is probably a numerical consequence of the particular value
  chosen for $v_A^c$. However, the exact strength of these shocks is
  not of crucial importance since they are mostly a consequence of the
  unphysical periodic boundary conditions, which serve to illustrate
  the effect of trapping the toroidal net flux on the structure of the corona.

\subsection{A thicker disc -- model S5}
\label{modelS5}

We now present the results of run S5 whose parameters are
described in table~\ref{model properties}.
This run used a thicker disc with $H/R=0.1$, and a correspondingly
smaller number of grid cells ($N_r$, $N_{\phi}$, $N_{\theta}$)=(360, 120, 210).
A distinct advantage of using a thicker disc is that in principle it
allows the use of a smaller number of grid cells while still giving rise to
a resolved model. We remind the reader that model S5 used
a ``viscous outflow boundary condition'' at the inner edge of the
computational domain (see section~\ref{setup} for details).

\subsubsection{Global disc properties}

\begin{figure}
\begin{center}
\includegraphics[scale=0.5]{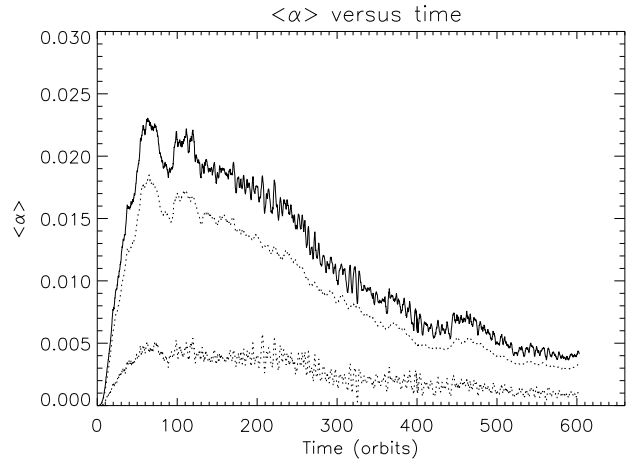}
\caption{Time history of the stress parameter 
$\langle \alpha \rangle$ for model S5.
The {\it dotted} line shows the contribution from the Maxwell stress,
the {\it dashed} line shows the contribution from the Reynolds stress,
and the {\it solid} line shows the sum of these. }
\label{alpha_t_S5}
\end{center}
\end{figure}

\begin{figure}
\begin{center}
\includegraphics[scale=0.5]{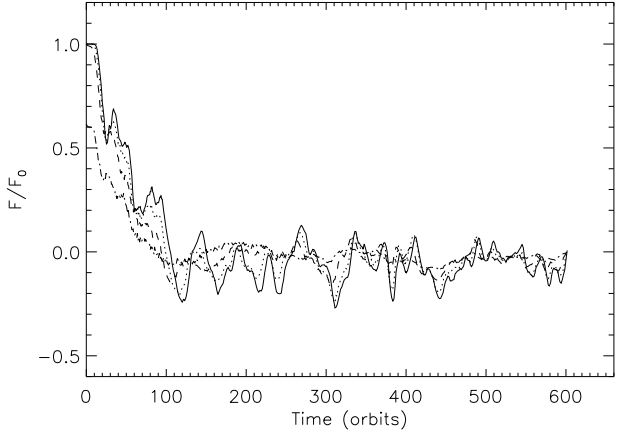}
\caption{Time history of the azimuthal magnetic flux for model S5.
The {\it dot-dashed} line shows the normalised magnetic flux contained within
$|\theta-\pi/2| \le H/R$ about the midplane, the {\it dashed} line shows
the flux within $|\theta-\pi/2| \le 2 H/R$, the {\it dotted} line
shows the flux within $|\theta-\pi/2| \le 3 H/R$ and the {\it solid} line
corresponds to the magnetic flux within the whole disc.
As for model S2, the initial flux within the disc escapes through the upper and
lower disc surfaces leading to a substantial reduction throughout the
disc. In particular, the regions around the midplane tend towards a state of
zero net flux. Oscillations in the total magnetic flux are caused by
magnetic flux entering through the vertical boundaries.}
\label{magflux_S5}
\end{center}
\end{figure}
The early evolution of this model proceeds very much along the lines
discussed in the opening paragraph of section~\ref{results}.
The time evolution of the volume averaged 
viscous stress parameter $\langle \alpha \rangle$ is presented in
figure~\ref{alpha_t_S5}, and shows that it saturates at a value
of $\langle \alpha \rangle \simeq 4 \times 10^{-3}$ 
after 500 orbits, in basic agreement with
the models S2, S3, and S4.
The time evolution of the azimuthal magnetic flux defined by
equation~(\ref{flux}) is shown in figure~\ref{magflux_S5}.
Similar behaviour is seen in model S5 as was described for run S2,
with magnetic buoyancy causing the initial flux to rise vertically
through the
disc into the corona where it escapes through the open boundary.
Close inspection of the {\it dot-dashed} line in figure~\ref{magflux_S5}
shows that the disc near the midplane contains essentially zero net magnetic
flux after about 100 orbits, such that sustained MHD turbulence there requires
the action of a dynamo. As already described, this feature of these 
simulations adds to the computational burden as zero net flux
simulations require high resolution in order that numerical
resistivity does not quench the MRI. The {\it solid} line in 
figure~\ref{magflux_S5} shows the total flux in the whole computational domain.
The fact that it oscillates about the zero line indicates that
the adopted boundary conditions lead to magnetic flux entering
the computational domain. The relatively small amount of flux that enters
suggests that this feature does not dramatically alter 
the results of our simulations.

\begin{figure*}
\includegraphics[scale=1.2]{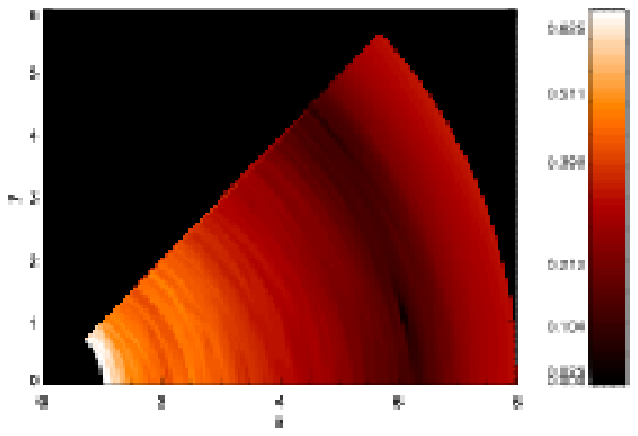}
\includegraphics[scale=1.2]{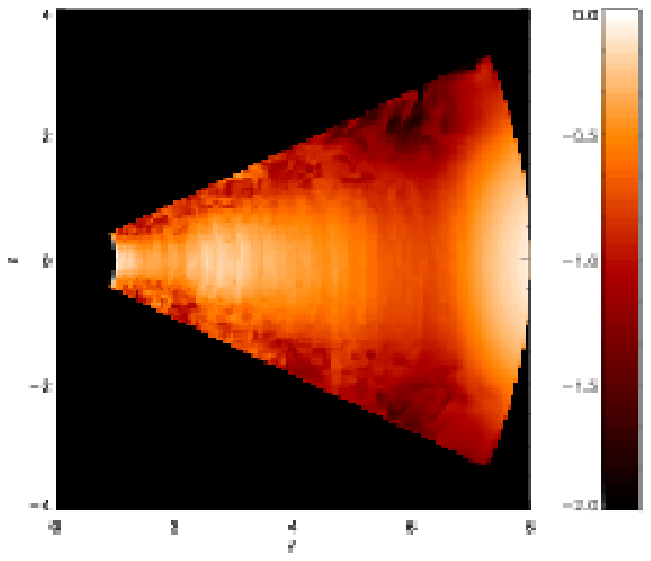}
\caption{The left panel shows an image of the density 
at the disc midplane after 500 orbits for model S5.
The right hand panel is a slice 
plotted in the $(r, \theta)$  plane showing the 
vertical density profile (this panel
shows the logarithm of the density after performing a transformation of
$\rho$ similar to that described in the caption of
figure~\ref{snapshots}). These images show that the action of
the stresses have caused signification mass transport within the disc,
resulting in an apparent density depression at $r \sim 6$.}
\label{density_S5}
\end{figure*}

Figure~\ref{density_S5} shows snapshots of the density in the disc
after 500 orbits. The left panel shows the density in the disc midplane,
and the right panel shows a vertical slice through the ($r$, $\theta$) plane.
The radial transport of mass during the simulation has caused 
a shallow depression to
form in the density profiles at radii between $5 \le r \le 7$ 
which are apparent in the figure. Also apparent are the trailing
spiral waves excited by the turbulence. These propagate radially through the
model and, because the disc is isothermal in the vertical
direction, these waves propagate with little vertical structure.
Animations of the disc density projected onto the ($r$, $\theta$)
plane indicate that individual prominent spiral waves which propagate radially
occupy most of the vertical extent of the disc.
Comparison between figures~\ref{density_S5} and \ref{snapshots}
shows that the `viscous outflow' boundary condition at the inner
edge of the disc is having the desired effect of preventing
a substantial build--up of mass there. As we comment in section~\ref{mass_flow},
some build--up of mass near the inner boundary does occur in this
model because the chosen value of the outflow velocity was
too small, but this should be a simple problem to remedy in
future models.

\begin{figure}
\begin{center}
\includegraphics[scale=1.2]{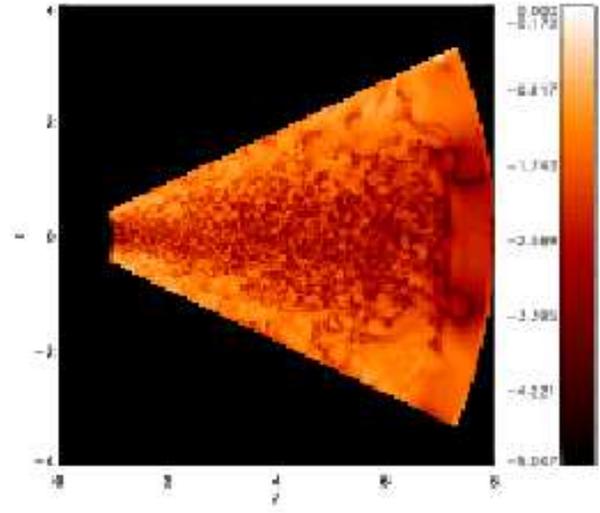}
\caption{This figure shows the logarithm of the Alfv\'en speed in
the disc after 500 orbits obtained by plotting a slice in 
the $(r,\theta)$ plane.}
\label{alfven_S5}
\end{center}
\end{figure}

Figure~\ref{alfven_S5} presents a vertical slice projected onto the
($r$, $\theta$) plane showing the logarithm of the Alfv\'en speed.
As observed in the runs S2 and S4, the Alfv\'en speed 
noticeably increases as one moves
away from the midplane to the disc surface above a height of about $2H$.
It is also clear from this figure that the disc is magnetically active all
the way down to the inner radial boundary due to the
`viscous outflow' condition used in this model.
The increase in relative strength of the magnetic
field with height is also illustrated by figure~\ref{mag_press} where the ratio of
magnetic pressure to gas pressure is plotted as a function of $\theta$
at radius $r=3.5$ using the {\it dot--dashed} line.
In agreement with the results of model S2, one sees
that $P_{mag}/P \simeq 0.01$ within $|\theta-\pi/2| < 2 H/R$, but it
quickly rises to $P_{mag}/P \simeq 0.5$ in the low density regions
above the midplane.

\begin{figure}
\begin{center}
\includegraphics[scale=0.5]{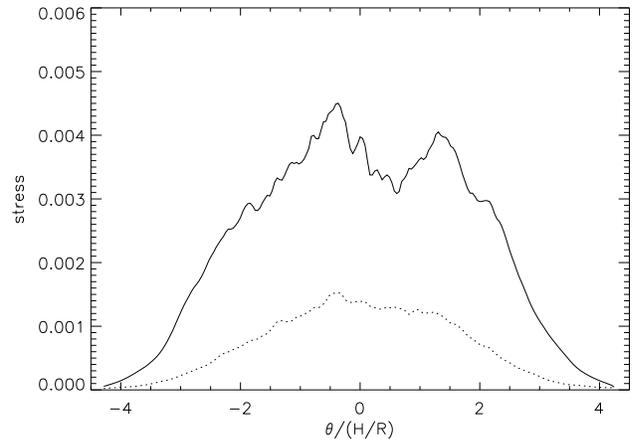}
\caption{The time averaged values of the Maxwell stress ({\it solid} line)
and Reynolds stress ({\it dotted} line), normalised to the midplane pressure,
as a function of $\theta$ for model S5. This was obtained by time averaging
for 100 orbits at radius $r=3.5$. This figure is similar to that obtained
for model S2, showing a similar drop-off of stress with height.}
\label{stress-v-theta_S5}
\end{center}
\end{figure}

The variation of stress with height is shown in
figure~\ref{stress-v-theta_S5}, which shows a similar fall--off in
the Maxwell and Reynolds stresses with height as observed for model
S2. This figure corresponds to radius $r=3.5$, and was obtained by
time averaging the stresses for 100 orbits.

\begin{figure}
\begin{center}
\includegraphics[scale=0.5]{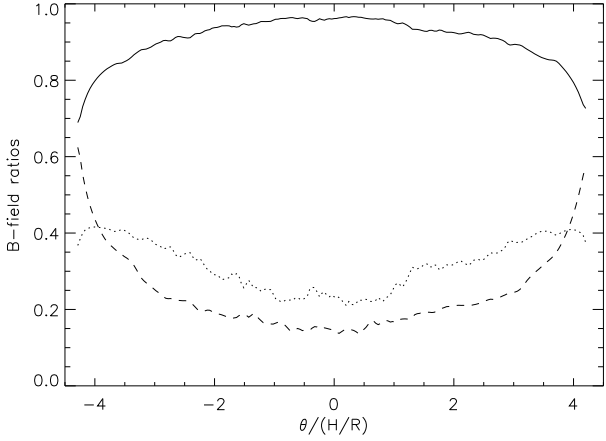}
\caption{Same as figure \ref{bfield_ratio_s2}. As in model S2, the
  azimuthal component remains  dominant throughout the vertical
  domain, but the field topology changes near the disc surface where
  the $\theta$ and $r$ components become larger.}
\label{B-ratio_S5}
\end{center}
\end{figure}

Figure~\ref{B-ratio_S5}, which should be compared with
figure~\ref{bfield_ratio_s2},
shows the variation of $|B_r|/B$,  $|B_{\phi}|/B$, and
$|B_{\theta}|/B$ versus $\theta$. This plot corresponds
to radial location $r=3.5$ and was obtained by time averaging for 100
orbits between $t=500$ -- 600 orbits. It shows a bigger increase of
the $r$ and $\theta$ component of the field in the upper layers of the
disc than in model S2. This is clearly influenced by the
boundary conditions imposed at the disc surface, where the field
is defined to be normal to the boundary with magnitude defined by
the $\nabla . \bb{B}=0$ condition in the NIRVANA runs
\citep[e.g.][]{hawley00}, but, as
suggested by the results of model S2 (which show a similar 
but reduced effect with different boundary conditions), 
it is probably also due to the
vertical stretching of field lines as localised regions of magnetised
fluid rise up from the disc midplane. 

\subsubsection{Velocity and density fluctuations}
\begin{figure}
\begin{center}
\includegraphics[scale=1.2]{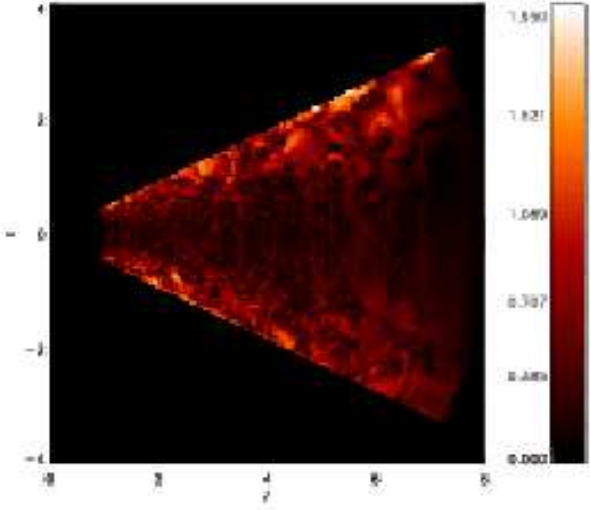}
\caption{Snapshots of the velocity fluctuations normalised by the
  local sound speed obtained in model S5. The results agrees very well
  with those of model S2 (see figure~\ref{mac_number})}
\label{dv_S5}
\end{center}
\end{figure}

The ratios of the velocity fluctuations in the $r$, $\theta$, and $\phi$
directions to the local sound speed were calculated and
found to be in very good agreement with the results of run S2
plotted in figure~\ref{vel_fluc_vert}.
The increase of the velocity perturbation
Mach number as one moves from the midplane to the disc surface
results in weak shocks
being generated in the corona, as illustrated
by figure~\ref{dv_S5} which shows a snapshot in the ($r$, $\theta$) plane
of the perturbed velocity divided by the sound speed. As observed in model
S2, and unlike model S4, typical Mach numbers for these shocks range between
1 -- 3. The fact that model S4 showed much stronger shocks illustrates
the role that magnetic forces in the corona 
have in exciting these supersonic motions.

The distribution of the density fluctuations in model S5 is represented in
figure~\ref{dens_fluc} with the {\it dashed line}. Simulation data were
averaged in time between $500$ and $600$ orbits, and in space in the
radial interval $3<r<5$ and in the height interval
$|\theta-\pi/2|<H/R$ to obtain this curve. Once again, the results are
in very good agreement with those of model S2.

\subsubsection{Mass flow in the disc}
\label{mass_flow}
Once model S5 had completed just over 600 orbits it was stopped.
It was restarted again at the 500 orbit mark, but with its
density field and azimuthal velocity reset to the values they
had initially at time $t=0$. All other variables (e.g. magnetic field,
vertical and radial velocities) had the values corresponding to the
500 orbit mark of the S5 run. This procedure was undertaken to smooth out
the large variations in surface density that arise during the 
early stages of these simulations because the stresses have 
large radial and temporal variations during these times.
The aim is to generate a turbulent protoplanetary disc with simple
surface density structure as a function of radius.

Having restarted the model it was run for 55 orbits, at which point
time averaging of the state variables and stresses within the disc were 
commenced. The simulation was continued for a further 100 orbits
while the time averaging was performed.

\begin{figure}
\begin{center}
\includegraphics[scale=0.5]{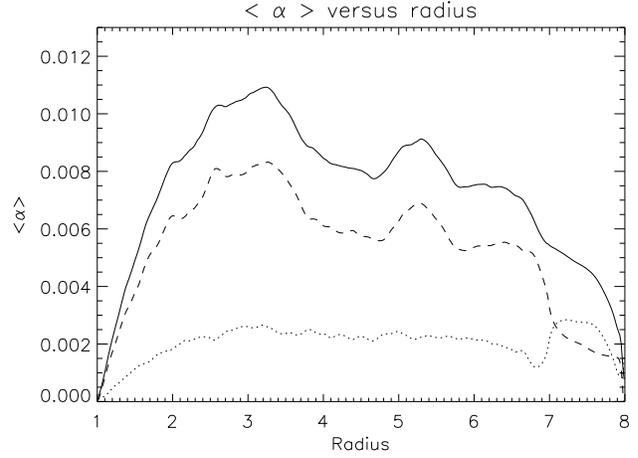}
\caption{Radial variation of the time averaged value of $\alpha$
for model S5. The curves were obtained by time averaging between
500--600 orbits. The {\it dashed} line corresponds to the
Maxwell stress, the {\it dotted} line to the Reynolds stress, and
the {\it solid} line to the sum of these.}
\label{alpha_v_r_S5}
\end{center}
\end{figure}

The radial variation of the time averaged $\alpha$ values in the disc are
shown in figure~\ref{alpha_v_r_S5}.
The results are similar to those obtained in model S2.
We note here, however, that the value of $\alpha$ is small 
($\sim 10^{-4}$) near the disc inner radial boundary
where a `viscous outflow' condition is imposed on the radial velocity.
This arises because during the initial phases of the run for
model S5, the magnetic field was non zero only in the
range $2.5 \le r \le 6$. As turbulence develops, mass
and magnetic field are transported inward, but at a rate
which is larger than accounted for by the `viscous outflow' condition.
The density in the midplane thus increases in this region, such that the
unstable modes of the MRI remain small and unresolved here 
[see equation~(\ref{crit})].
Consequently turbulence in the inner most region of the disc
remains weak. A remedy for this would be to modify the
`viscous outflow' condition so that it responds more accurately
to the inflow of matter from further out rather than using 
a prescribed inward velocity as was done for model S5.

\begin{figure}
\begin{center}
\includegraphics[scale=0.5]{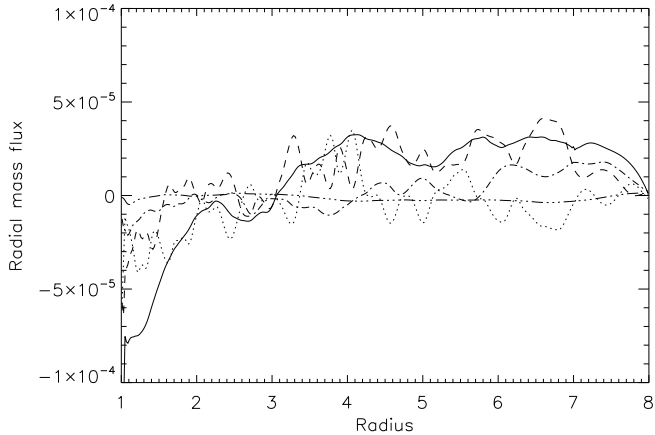}
\caption{Radial mass flux versus radius for model S5.
The {\it dotted} line corresponds to $| \theta - \pi/2| \le H/R$,
the {\it dashed} line corresponds to $H/R <  |\theta - \pi/2| \le 2 H/R$,
the {\it dot-dashed} line corresponds to $2H/R <  |\theta - \pi/2| \le 3 H/R$,
and the {\it dot-dot-dot-dashed} line corresponds to 
$3H/R <  |\theta - \pi/2| \le$ disc surface. The {\it solid} line
represents the total radial mass flux through the disc.}
\label{radmflux_S5}
\end{center}
\end{figure}

Figure~\ref{radmflux_S5} shows the time averaged radial
mass flux through the disc as a function of radius.
Each line corresponds to mass flow at a different height in the disc.
The {\it dotted} line corresponds to the region within $\Delta \theta \le H/R$
about the midplane, where $\Delta \theta = |\theta-\pi/2|$.
The {\it dashed} line corresponds to the
region bounded by $H/R < \Delta \theta \le 2 H/R$, the {\it dot--dashed}
corresponds to $2H/R < \Delta \theta \le 3 H/R$, and the 
{\it dot-dot-dot-dashed} line corresponds to  
$3H/R < \Delta \theta \le (\theta_{max} \,\, {\rm or} \,\, \theta_{min})$.
The total radial mass flux is shown by the {\it solid} line (and is the sum
of all the other lines).
Evidently negligible mass is transported in the upper most parts
of the disc corona, as there is very little mass there.
The other three regions considered, however, all 
contribute significantly to the mass flux. In the outer regions of the disc
the zone near the midplane appears to be transporting mass inward
whereas the upper regions of the disc are transporting it
outward, suggesting that the long term mass flow in 
vertically stratified turbulent discs can be a complicated function of
disc height. A similar picture was described by \citet{devilliers&hawley03}
for simulations of accretion tori around black holes.

\begin{figure}
\begin{center}
\includegraphics[scale=0.5]{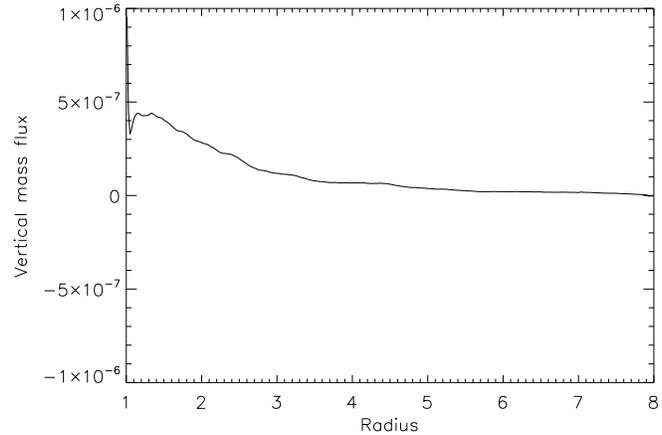}
\caption{The sum of the mass fluxes through the vertical boundary of the disc
located at $\theta=\theta_{max}$ and $\theta_{min}$.}
\label{thetamflux_S5}
\end{center}
\end{figure}

The total vertical mass flux through both boundaries at the disc surface
is shown in figure~\ref{thetamflux_S5}. 
It is clear that the disc drives a vertical mass flow at a rate
that is less than two orders of magnitude below that which occurs
radially in the disc. A similar result was obtained in model S2.
However, the restricted
size of our meridional domain prevents us from commenting in detail
about any wind that may be launched from the disc surface.

\begin{figure}
\begin{center}
\includegraphics[scale=0.5]{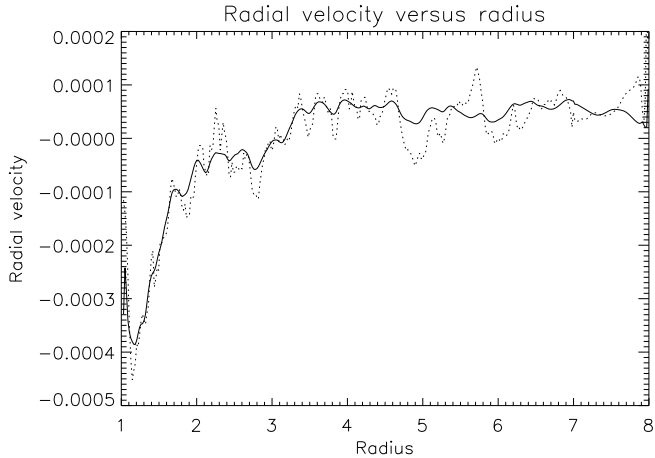}
\caption{Comparison between the time averaged radial velocity obtained
in the simulations ({\it solid} line) and the predicted value of
$v_R$ obtained from equation~(\ref{v_R}) shown by the {\it dotted} line.}
\label{vr_v_r_S5}
\end{center}
\end{figure}

We now turn to the question of how well the averaged
radial velocity in the disc agrees with the expectations of
thin disc theory as described by equation~(\ref{v_R}). We computed
the time average of $\Sigma(R)$, ${\overline T_M}(R)$, ${\overline T_R}(R)$
and using a simple finite difference approximation calculated
the expected radial velocity profile ${\overline v_R}$ in the disc.
The results are shown using the {\it dotted} line in figure~(\ref{vr_v_r_S5}),
where the actual value of ${\overline v_R}$ obtained in the simulation is shown
using the {\it solid} line. Although the effect of fluctuations remain,
the agreement between the predicted and actual values is remarkably good.
This demonstrates that the vertically stratified turbulent
discs considered here behave very much like standard $\alpha$ discs
when their evolution is considered over long time scales.
On short time scales, however, the differences are self--evident.
We note that a comparison of the predicted and actual values of
${\overline v}_R$ was undertaken for model S4
and gave rise to a very similar level of agreement.

\section{Conclusions}
\label{conclusion}
In this paper we have presented the results of 3-D MHD simulations of
stratified and turbulent protoplanetary disc models. Our primary 
motivation is to develop disc models that can be used to examine
outstanding issues in planet formation such as the migration of
protoplanets, the growth and settling of dust grains, gap formation
and gas accretion by giant planets, and the evolution and influence
of dead--zones. Given that these phenomena occur on secular time scales,
a key requirement is the development of disc models which are able to
achieve a statistical steady state and sustain turbulence over long 
run times.

We examined the issue of numerical resolution, and found that
disc models with $\simeq 15$ vertical zones per scale height in the
vertical direction 
showed a continuing slow decline in their magnetic activity, 
and gave rise to relatively small values of $\langle \alpha \rangle$. 
A suite of higher resolution runs with $\simeq 25$ zones
per scale height achieved statistical steady states with
values of $\langle \alpha \rangle \simeq 4 \times 10^{-3}$, and it was shown
that these models resolve the fastest growing modes of the MRI
throughout the disc once a turbulent steady state has been achieved.
For this reason we focused on simulations performed using this
higher resolution.
\noindent The key features of the resulting disc models are:
\begin{itemize}
\item{Any toroidal magnetic flux that is initially present within
the disc is quickly expelled from the midplane due to
magnetic buoyancy. This occurs on a time scale of $\sim 100$ orbits,
which is the time required for the MRI to grow and
develop into non linear turbulence throughout the disc. 
The disc then evolves as
if it is threaded by an approximately zero net flux magnetic field, 
such that high resolution is required to maintain turbulent activity.}

\item{A quasi--steady state turbulent disc is obtained after run times
of between 250 -- 500 orbits, depending on the model.
The volume averaged value of the
effective viscous stress paramater 
$\langle \alpha \rangle \simeq 4 \times 10^{-3}$,
and time averaged radial profiles of $\alpha$ yield variations
of no more than a factor of two within the active domains of
the disc models. These results are in basic agreement with
previous studies of cylindrical discs \citep{hawley01,pap&nelson03a}}

\item{The discs can be described as having a two--phase global structure
as a function of height:
a dense, magnetically subdominant turbulent core that is unstable to the MRI
in regions within $|Z| < 2.5 H$ of the midplane, above and
below which exists a highly dynamic and magnetically dominated corona 
which supports weak shocks and is stable against the MRI.
The engine that drives this structure is the MRI which
generates and amplifies magnetic field near the midplane, which
then buoyantly rises up into the low density corona where
it dissipates and flows out through the boundaries at the disc surface.
This is in basic agreement with the shearing box simulations
presented by \citet{miller&stone00} and 
recent studies of thick tori orbiting around black holes
\citep{hawley00,hawley&krolic01,hawleyetal01,devilliers03}}. 

\item{The velocity and density fluctuations generated by the models
were found to be smaller than those obtained in cylindrical disc
simulations using toroidal net flux magnetic field configurations 
\citep{nelson05,fromang&nelson05}. $\delta \rho/\rho_0 \simeq 0.08$
in the stratified runs whereas in the cylindrical disc runs with net
flux 
$\delta \rho/\rho_0 \simeq 0.13$. This has implications for the 
dynamics of dust, planetesimals and low mass protoplanets
in turbulent discs as their stochastic evolution is driven by
these fluctuating quantities.}

\item{The vertically, azimuthally and time averaged values of the radial
velocity in some of the disc models were compared
with the expectations of viscous disc theory, and were
found to give very good agreement. We conclude that, 
subject to a suitable
time average, global evolution of these stratified turbulent
models is in good accord with standard viscous disc theory
\citep[e.g.][]{shakura&sunyaev73,balbus&pap99,pap&nelson03a}}
\end{itemize}

There are a number of outstanding issues raised by our simulation results.
\citet{fromang&nelson05} reported the presence of anticyclonic
vortices in turbulent unstratified
cylindrical disc models. In the stratified models
we present in this paper, however, we did not find any evidence of
vortices. This could be due to a number of reasons. First,
the vertical stratification may prevent the formation of vortices
near the midplane. A study by \citet{barranco&marcus05}
showed that column--vortices in stratified discs are unstable
and are quickly destroyed. They showed that vortices could form
in the more strongly stratified upper regions of the disc
which are more than one scale height above the midplane.
In our simulations these regions are typically dominated
by magnetic fields, whose associated stresses
may prevent the formation
of vortices there. Second, we adopted 
a smaller azimuthal domain than \citet{fromang&nelson05}: $\pi/4$ versus
$\pi/2$ and $2 \pi$ models. 
This may prevent the formation of vortices, which were found to
be quite extended in $\phi$ by \citet{fromang&nelson05}.
We did not observe any vortices in the cylindrical disc model
C1 described in section~\ref{cyl_setup}, and this may be partly explained by the
smaller azimuthal domain. Finally, the different magnetic
field topology contained in the disc may play a role:
\citet{fromang&nelson05} used a net flux
toroidal magnetic field. In the stratified models, the toroidal flux is
quickly expelled from the disc midplane and the evolution is 
more similar to that of a zero net flux disc. The cylindrical model C1
contained a zero net flux magnetic field. The properties of the
field can affect the turbulence and hence the formation of vortices.
In particular, stronger spatial and temporal variations in the stresses
may cause the surface density variations to differ systematically
between the models presented here and those in \citet{fromang&nelson05}.
If the formation of vortices in the models described in
\citet{fromang&nelson05} are related to the `planet modes' described
by \citet{hawley87}, then these differences may explain the
lack of vortices seen in the stratified models and model C1.
These issues, and their influence on the evolution of solid
bodies will be explored in greater detail in a future paper.

Finally, this is the first paper in a series which
describes an approach to setting up   
models of turbulent, stratified protoplanetary discs capable
of sustaining turbulence over long run times.
Future papers will present a systematic study of outstanding
problems in planet formation theory, such as
disc--planet interactions, dust and planetesimal
dynamics, and effects related to the presence of a dead zone. We also
note that these models themselves can be further improved by including
a realistic equation of state, heating and cooling of the disc,
and a self--consistent treatment of the evolving ionisation fraction 
and conductivity of the disc material. At the present time inclusion
of these physical processes is beyond current computational resources.

\section*{ACKNOWLEDGMENTS}
The simulations presented in this paper were performed using the 
QMUL High Performance Computing Facility purchased under the SRIF 
initiative, and the UK Astrophysical Fluids Facility (UKAFF).
The research was funded by a PPARC research grant PP/C507501/1.

\bibliographystyle{aa}
\bibliography{author}

\end{document}